\documentclass{mn2e}

\usepackage{color}

\usepackage{graphicx}
\usepackage{amsbsy}
\usepackage{amsmath}
\usepackage{epsf}

\newcommand{\apj}{ApJ}
\newcommand{\apjl}{ApJL}
\newcommand{\mnras}{MNRAS}
\newcommand{\apjs}{ApJS}

\newcommand{\aap}{A\&A}

\newcommand{\beq}{\begin{equation}}
\newcommand{\eeq}{\end{equation}}

\DeclareMathAlphabet{\mathsfsl}{OT1}{cmss}{bx}{sl}
\SetMathAlphabet{\mathsfsl}{bold}{OT1}{cmss}{bx}{sl}

\usepackage{epstopdf}
\begin{document}

\title[$\rm^{13}CO$ Filaments in the Taurus Molecular Cloud]
{$\rm^{13}CO$ Filaments in the Taurus Molecular Cloud}
%

\author[G.~V. Panopoulou et al.]
  {G.~V.~Panopoulou$^{1,2}$, K.~Tassis$^{1,2}$, P.~F.~Goldsmith$^{3}$, M.~H.~Heyer$^{4}$ \\
    $^1$Department of Physics, University of Crete, PO Box 2208, 71003 Heraklion, Greece\\
    $^2$IESL, Foundation for Research and Technology-Hellas, PO Box 1527, 71110 Heraklion, Crete, Greece \\
    $^3$Jet Propulsion Laboratory, California Institute of Technology,
    4800 Oak Grove Blvd., Pasadena, CA 91109, USA \\
    $^4$Department of Astronomy, University of Massachusetts, Amherst, MA}

\maketitle 

\label{firstpage}
\begin{abstract}
We have carried out a search for filamentary structures in the Taurus molecular cloud using $\rm^{13}CO$ line emission data from the FCRAO survey of $\rm \sim100 \, deg^2$. We have used the topological analysis tool, DisPerSe, and post-processed its results to include a more strict definition of filaments that requires an aspect ratio of at least 3:1 and cross section intensity profiles peaked on the spine of the filament. In the velocity-integrated intensity map only 10 of the hundreds of filamentary structures identified by DisPerSe comply with our criteria. Unlike Herschel analyses, which find a characteristic width for filaments of $\rm \sim0.1 \, pc$, we find a much broader distribution of profile widths in our structures, with a peak at 0.4 pc. Furthermore, even if the identified filaments are cylindrical objects, their complicated velocity structure and velocity dispersions imply that they are probably gravitationally unbound. Analysis of velocity channel maps reveals the existence of hundreds of `velocity-coherent' filaments. The distribution of their widths is peaked at lower values (0.2 pc) while the fluctuation of their peak intensities is indicative of stochastic origin. These filaments are suppressed in the integrated intensity map due to the blending of diffuse emission from different velocities. Conversely, integration over velocities can cause filamentary structures to appear. Such apparent filaments can also be traced, using the same methodology, in simple simulated maps consisting of randomly placed cores. They have profile shapes similar to observed filaments and contain most of the simulated cores.

\end{abstract}


\section{Introduction}
\label{sec:intro}
Stars form in cold, dense, molecular clouds in the interstellar medium (ISM). The initial conditions of star formation remain unclear, so studies of the structure of the parent molecular cloud are of utmost importance. Recently, the Herschel Space Observatory has provided high sensitivity infrared images of the cold ISM with sub-parsec resolution. Analyses of column density maps derived from Herschel observations of nearby molecular clouds show the ubiquitous existence of elongated overdensity structures with aspect ratios (length to width) of $5 - 10$, referred to as \textit{filaments} \cite{andre2010,molinari2010}. Filaments have been studied extensively in both dust emission (Men'shchikov et al. 2010; Arzoumanian et al. 2011; Peretto et al. 2012; Palmeirim et al. 2013) and molecular line emission maps (Nagahama et al. 1998; Hacar \& Tafalla 2011; Li \& Goldsmith 2012; Henshaw et al. 2013; Arzoumanian et al. 2013; Hacar et al. 2013).

Those structures found in Herschel dust maps have mean radial density profiles that have a flat inner part surrounded by an outer envelope with density falling as $\propto r^{-2}$ \cite{arzoumanian2011}, similar to cores (Ward-Thompson et al. 1994; Andr\'{e}, Ward-Thompson, Motte 1996; Ward-Thompson, Motte \& Andr\'{e} 1999; Bacmann et al. 2000). Moreover, Herschel dust filaments appear to have a universal characteristic width of 0.1 $\pm$ 0.03 pc \cite{arzoumanian2011}.

The Herschel results on the ubiquity of filaments point towards a picture of star formation in which filaments may constitute an early stage of the process \cite{andre2013}. According to this scenario, filaments form first inside the parent molecular cloud, and then break up into cores, where ultimately protostars appear.

Herschel continuum emission data have the disadvantage that they contain no velocity information, and as such may connect in projection structures that are separated by significant distances in three dimensions. Such concerns may be addressed through studies of molecular line emission maps, which, in addition to velocity-integrated intensity, can also reveal separate line-of-sight velocity components. In the Taurus molecular cloud, Hacar et al. (2013) studied the line emission of $\rm C^{18}O, N_2H^+$ and $\rm SO$ in the well known filament L1495/B213. They found that this filament is comprised of a multitude of velocity-coherent sub-structures with typical lengths of 0.5 pc and aspect ratios larger than 3. They proposed that filaments may be bundles of intertwined velocity-coherent sub-filaments. Palmeirim et al. (2013) identified striations perpendicular to the L1495/B213 filament in both Herschel dust emission and $\rm^{13}CO$, suggesting the accretion of material through them onto the main structure.

The scope of this paper is to identify and study filamentary structures in the largest area of the Taurus molecular cloud studied to date, in molecular line emission. The data are provided by the Five College Radio Astronomy Observatory (FCRAO) survey of the Taurus cloud in $\rm^{13}CO$ $\rm J = 1-0$ emission \cite{narayanan2008}, presented in section \ref{sec:data}. A study of this size can allow statistical characterization of filaments in position-position-velocity space. First, we search for filaments in the entire map, integrated over the velocity range 0.25 km/s to 9.8 km/s. Integration of the emission over a large velocity range produces an image similar to those seen in dust emission. We use the DisPerSe algorithm \cite{sousbie2011} to trace linear structures and then apply an algorithm that we developed for discarding structures that are non-astrophysical (both algorithms are described in section \ref{sec:algorithms}). The analysis results in the identification of fewer than 10 filaments in the entire map (section \ref{sec:results}), whose velocity structures imply that (assuming a cylindrical shape) they would disperse in less than 1 Myr (section \ref{ssec:velocities}). The scarcity of filamentary structures leads to the consideration of projection effects which may affect the study and are discussed in section \ref{sec:effects}. Indeed, analysis of velocity channel maps reveals the existence of a multitude of filamentary structures (section \ref{sec:slices}). Finally, in section \ref{sec:discussion} we provide a summary of the basic conclusions of this study.

\section{Data}
\label{sec:data}
\begin{figure}
 \centering
\includegraphics[scale=1]{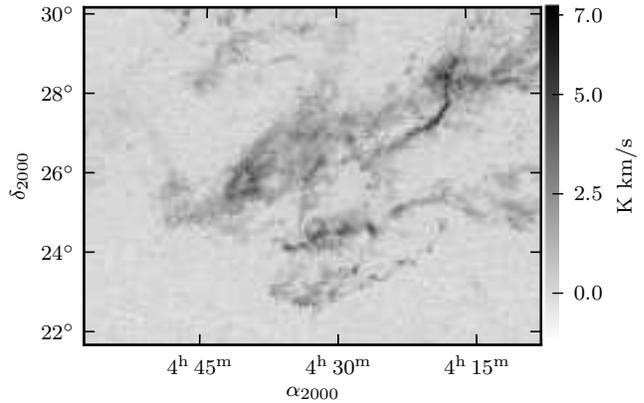}
 \caption{Integrated $\rm ^{13}CO$ emission map of the Taurus region integrated over the range 0.25 km/s to 9.8 km/s.}
  \label{fig:integrated13co}
    \end{figure}

In this work, we use $\rm ^{13}CO$ data in the form of a (x,y,v) cube of the Taurus molecular cloud as observed with the 13.7 m FCRAO telescope \cite{narayanan2008}. The map covers an area of $\rm 98\, deg^2$ corresponding to a region 28 pc by 21 pc. The FWHM beam width of the telescope is 47" at 110.2013541 GHz. The angular spacing (pixel size) of the re-sampled on the fly data is 20" \cite{goldsmith2008}, which corresponds to a physical scale of $\rm \approx 0.014$ pc at the distance of the Taurus cloud (140 pc). There are 76 velocity channels in the $\rm ^{13}CO$ data cube. The width of a velocity channel is $ \delta V_{ch}=0.266 \, \rm km/s$. A map of the region, where the emission has been integrated over the range 0.25 to 9.8 km/s, is shown in figure \ref{fig:integrated13co}.

\section{Data analysis and Filament identification}
\label{sec:algorithms}

\subsection{The DisPerSe software}
\label{ssec:Disperse}
We use the DisPerSe software \cite{sousbie2011} to identify the topology and consequently the filamentary structures in the $\rm ^{13}CO$ intensity maps. The DisPerSe (Discrete Persistent Structures Extractor) software is designed to extract the prominent edges of a density field, referred to as \textit{filamentary structures}. Its main application has been to trace the web-like pattern of galaxies (cosmic web) both in cosmological simulations and redshift catalogs. However, it has also been widely used for filamentary structure extraction from Herschel images in recent publications \cite{arzoumanian2011,peretto2012,palmeirim2013}.

DisPerSe extracts the topological skeleton\footnote{The topological skeleton of a shape is the smallest possible set of lines that are equidistant to the borders and preserve the topology.} of an image using a method based on discrete Morse Theory. Thinking of an intensity map as a terrain comprised of peaks (local maxima), valleys (around local minima) and saddle points, a skeleton is made up of the ridges (filamentary structures) that connect peaks and saddle points. Discrete Morse theory can not provide a measure of how significant topological features are (for example noise can produce spurious features). Therefore, a simplification that removes less significant topological features is necessary to determine the dominant structures within the data. DisPerSe uses the concepts of \textit{persistence} and \textit{robustness} for this process. Persistence represents the absolute difference of the intensity value between two critical points of different kinds (e.g. a maximum and a minimum). Setting an appropriate persistence threshold is a way to filter noise and non-meaningful structures. Robustness is a measure of how contrasted filaments as a whole are with respect to their background \cite{weinkauf}.

The DisPerSe software uses these mathematical tools to extract structures from intensity/density maps. Skeleton files can be post-processed by DisPerSe in several ways. Options include smoothing over a number of pixels, assembling arcs and trimming for robustness.
It is important to note that there is \textit{no absolute measure of the `goodness' of a skeleton}. The DisPerSe parameters provide an essential, yet non unique tool for evaluating the accuracy of the algorithm's result. It is clear that the introduction of the persistence and robustness thresholds aids the proper detection of significant features. However, in the case of an intensity map with structures of various emission strengths, a selection of high robustness causes true features in less intense regions to be discarded.
Given the lack of an established, rigorous method, the search for an appropriate skeleton is largely subjective. In this project, we choose to resolve this issue by performing a parameter study of different combinations of persistence, robustness and assembling arcs (which are the main parameters) for all of our images. 

\subsection{Extracting properties of structures within the DisPerSe skeletons}
\label{ssec:profile-filtering}

The DisPerSe software identifies elongated intensity structures in images. Having a mathematically accurate map of these structures is of paramount importance for studies such as our own. With this information in hand, it is possible to investigate the properties of the structures in the Taurus molecular cloud, such as width, length and kinematics. 

Analysis of the direct output of DisPerSe has shown, however, that the `filaments' found by this algorithm are not always acceptable astrophysical filaments but may exhibit a number of undesirable properties (e.g. breaks). Thus, we developed a secondary analysis to filter the output skeleton and redefine its structures. The algorithms that do this are described in sections \ref{sec:profiling_alg} and \ref{sec:redefining_bones}. In order to precisely define what is meant by the term `filament', we adopt a different terminology than that used in the DisPerSe software. We refer to arcs comprising the skeleton as \emph{bones} and we use the term \textit{filaments} only for structures that survive our algorithm quality checks. These require that 
\begin{itemize}
\item the intensity profiles of cuts perpendicular to the structure are centrally peaked 
\item the structure is continuous and discernible from background diffuse emission
\item filaments have an aspect ratio (length to width ratio) typical of an elongated form.
\end{itemize}

\subsection{Profile fitting and filtering algorithm}
\label{sec:profiling_alg}

Interesting insight into the structure of a bone can be gained by looking at its cross sections (or 1 dimensional radial intensity profiles) at different points along its length. The two significant properties of such a profile are its shape and width. It has been found in the Herschel dust continuum data that the form of the column density profile of filaments follows a Plummer function \cite{arzoumanian2011}
\begin{equation}
\centering
 \Sigma_p(r) = \Sigma_c \frac{\rho_c R_{flat}}{[1+(r/R_{flat})^2]^{\frac{p-1}{2}}},
\label{eqn:plummercolumn}
\end{equation}
with p =2, where r is the apparent distance from the axis of the filament on the plane of the sky, $\rm \rho_c$ is the central density, $\Sigma_c$ the central column density and $\rm R_{flat}$ the characteristic radius of the inner flat portion of the profile \cite{plummer1911}. 

For the determination of the width, a simplistic approach such as that of finding the radius at which the intensity has reached half maximum can fail under some conditions. For example, the profile shown in figure \ref{fig:halfmaxprof} has no point on it that coincides with the half maximum value. It follows from examples like this, that a way to model the structure's cross section that is independent of its neighboring structures is needed.
\begin{figure}
 \centering
\includegraphics[scale=1]{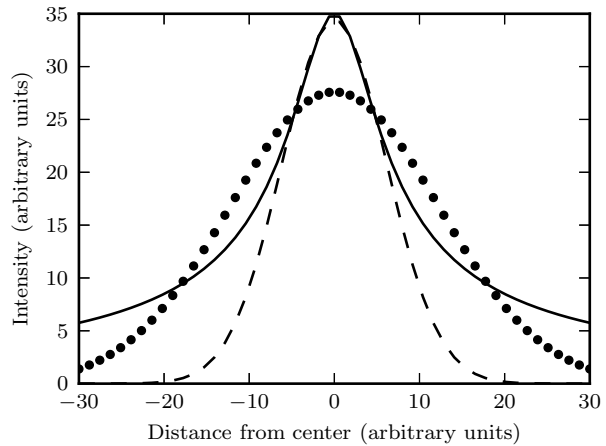}
 \caption{Plummer function with $\rm R_{flat} = 5$ and $\rm p = 2$ (solid line), and fitted Gaussians. Dotted Gaussian is fitted to the entire range of radii whereas the dashed Gaussian is fitted only to the central points. $\rm FWHM_{dashed} \approx 3 R_{flat}$ but $\rm FWHM_{dotted} \approx 6 R_{flat}$.}
  \label{fig:plummer-gauss}
    \end{figure}

\begin{figure}
 \centering
\includegraphics[scale=1]{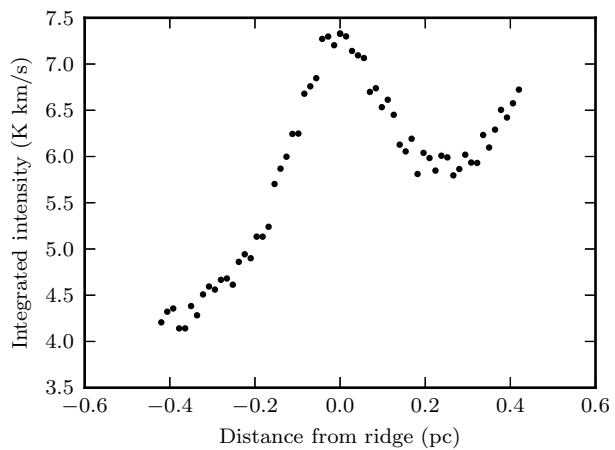}
 \caption{Profile with no point at half maximum value.}
 \label{fig:halfmaxprof}
\end{figure}
Many studies, including the present one, choose to define the width of a profile by fitting a Gaussian function to it. Arzoumanian et al. (2011) found that the FWHM relates directly to $\rm R_{flat}$ as 
\begin{equation}
\centering
\rm{FWHM} \approx 3 R_{flat}.
\label{eqn:fwhmrflat}
\end{equation}
A graphical representation of this can be seen in figure \ref{fig:plummer-gauss}, where the dashed line is a Gaussian with FWHM equal to three times the $\rm R_{flat}$ of the Plummer function (solid line). This ideal fit is accomplished by fitting only the inner part of the profile. A Gaussian fit to the entire profile, such as the dotted line, tends to overestimate the width of the profile by a factor of 2.

A standard approach is to study the mean radial profile of the structure (the average of all profiles along it). In this study, however, we choose to fit all profiles along a bone individually, in addition to the commonly used mean profile, so as to examine the variation of the width along the filament. The algorithm `cuts' bones into densely spaced cross sections and fits Gaussians to them. First, a skeleton found by DisPerSe is provided as input. Each structure in the skeleton is defined by a set of points along it, called \textit{sampling points}. Each bone is divided into a list of line segments connecting its sampling points. At each sampling point, the segment connecting it to the previous one is used to find a perpendicular direction. Pixels in this direction comprise the intensity profile of the structure at that sampling point. A Gaussian function is fitted, by using the function optimize.leastsq of the python library SciPy \cite{scipy}. 

The extent of the region over which we carry out a fit to the profile affects the fitted function's parameters significantly. More specifically, we can not use a simple Gaussian fit to each profile, because as shown in figure \ref{fig:plummer-gauss}, a Gaussian fitted to the entire profile tends to miss the central part completely. A fit must only be applied to the central part of the profile. As we do not know the width of the filaments present in our data a priori, we have to create an approach that handles this lack of knowledge and the diversity of the structures as automatically and well as possible. Therefore, the algorithm performs what we call `dynamical fitting'. Beginning from a constant given profile extent (2R), it fits Gaussians recursively, to smaller and smaller distances from the central point, removing two pixels on each side of the profile at each step. It stops at a given extent that we assume is the minimum that can determine the bone widths (20 pixels for our data, corresponding to 0.28 pc). 

\begin{figure}
 \centering
\includegraphics[scale=1]{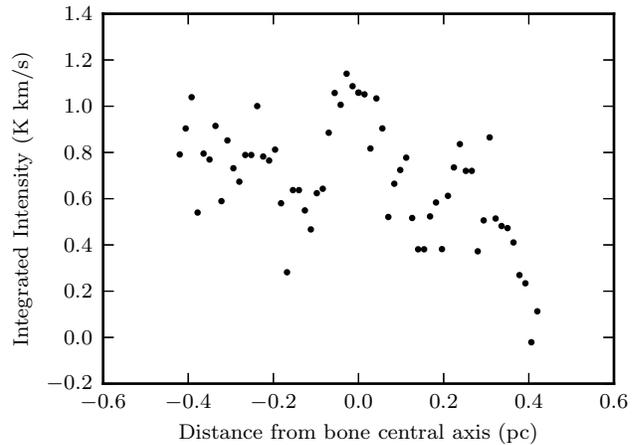}
\caption{Example of profile buried within the noise level.}
  \label{fig:profs-bad}
    \end{figure}

\begin{figure}
\includegraphics[scale=1]{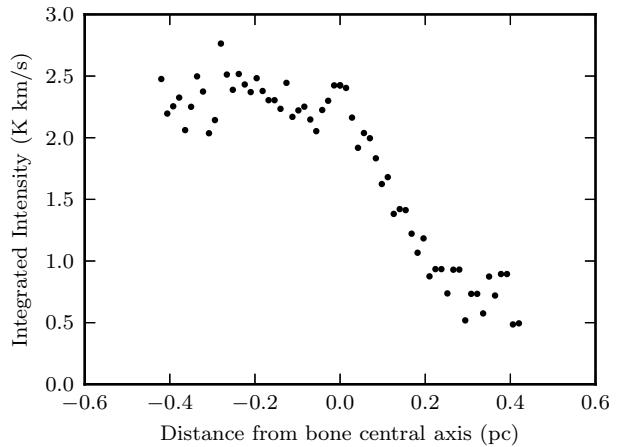}
\caption{Example of skeleton profile on the side of a real structure.}
  \label{fig:profs-bad2}
    \end{figure}

\begin{figure}
 \centering
\includegraphics[scale=1]{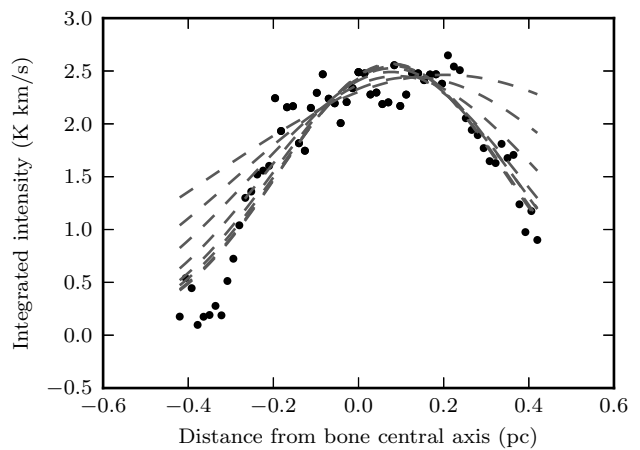}
\caption{Example of an acceptable profile. Gray lines are Gaussian fits resulting from the dynamical fitting process.}
\label{fig:profs-good}
\end{figure}

In this way, we obtain about twenty fits for a certain profile. The result of the dynamical fitting process is used to test whether a profile is `acceptable'. If the profile is smoothly peaked, the fits should be clustered around the shape of the profile. This means that the mode of the FWHM distribution of the fits will be the one closest to reality. Otherwise, if the profile is one-sided, flat etc, the frequency of the mode should be 1, meaning no fit looks like the others. Examples of profiles with these undesirable shapes are shown in figures \ref{fig:profs-bad} and \ref{fig:profs-bad2}. The profile shown in figure \ref{fig:profs-bad} is buried within the noise level at 1 K$\cdot\rm km/s$, while that in figure \ref{fig:profs-bad2} is dominated by a neighboring structure. Such examples are not uncommon in DisPerSe skeletons. Consequently, it is essential to establish a set of criteria that can allow us to automatically distinguish between profiles that can be accepted as parts of real structures and those that cannot. An example of a profile that can be accepted as part of a real filamentary structure is shown in figure \ref{fig:profs-good}. We can not trust standard goodness of fit measures, such as the reduced chi-squared, to determine the acceptability of a profile, because of the peculiarity of profile shapes.


The calculation of the mode frequency alone is not always sufficient for the distinction of a profile as acceptable. There are many cases where the profile is in no way similar to a smooth peak, but the fits converge for a small number of iterations.  

\begin{figure*}
 \centering
\includegraphics[scale=1]{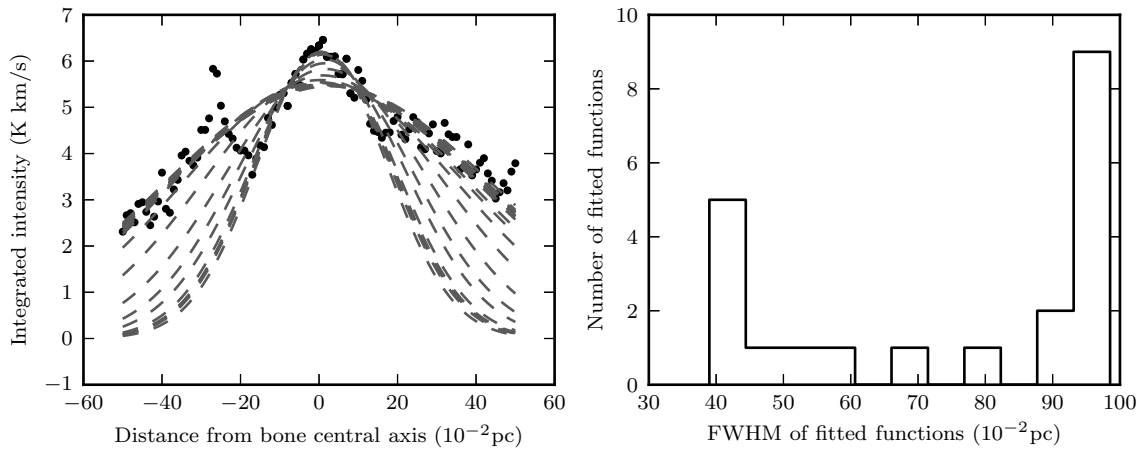}
\caption{Dynamical fitting of a profile with bimodal distribution of FWHMs. Left: Profile points depicted as dots; the dashed lines are the Gaussian fits that result from the process of dynamical fitting. Right: Distribution of FWHM returned from dynamical fitting.}
\label{fig:profs-bimodal}
\end{figure*}

The criteria used to determine the quality of the profile are:
\begin{itemize}
\item The frequency of the mode of the distribution of FWHMs must be larger than unity, i.e. the profile must have at least one peak.
\item The fit corresponding to the mode of FWHMs must have a mean within 5 pixels of the bone's central axis, i.e. the profile is centrally peaked.
\item The difference between the peak intensity of the last fit and the intensity value of the lowest point in the profile must be larger than the noise level. 
\end{itemize}
The background (or noise) level is defined as 5 times the noise level estimated from Goldsmith et al. (2008), section 2. For $\rm^{13}CO$, the mean rms antenna temperature is equal to 0.125 K in channels of 0.27 $\rm km/s$ width.

Finally, there are cases of bimodal distributions of FWHM such as that in figure \ref{fig:profs-bimodal} (left). Data points are depicted as dots and dashed lines are the Gaussian fits that result from the process of dynamical fitting. In this case, the wider FWHM will be assigned to the particular profile. In similar cases of multi-peaked FWHM distributions, if the principal mode of the distribution corresponds to a fit that is not centrally peaked, then the profile will be wrongly discarded. Therefore, we give profiles not satisfying the three criteria one more chance. The profile is set to the test once more but with the second maximum of the FWHM distribution regarded as the correct one. In the particular case shown in figure \ref{fig:profs-bimodal}, that would be the value at 0.4 pc shown in the profile's distribution of FWHM (figure \ref{fig:profs-bimodal} right). All profiles that satisfy the above three conditions (on the first or the second chance) are flagged as `acceptable' and the mode of FWHMs (of every profile) is assigned as the width of the bone at that cross section. 

\subsection{Redefining bones}
\label{sec:redefining_bones}
The next step is to use the information obtained by the profiling process to redefine the bone structures. When used conservatively, in order to increase the length of the structures in a skeleton, DisPerSe may connect structures that are not necessarily a whole (see section \ref{sec:effects}). Therefore, we need a way to distinguish which bones are really an entity and which are not. The question is, how does one define whether something is an entity? We choose this definition: A bone is a single entity if along its entire central axis, its cross sections are what we defined to be acceptable. So if along a bone there are gaps (e.g. profiles there were not peaked) the bone is not a single structure.  

We use another algorithm to redefine bones according to this definition. We use a tolerance for unacceptable consecutive profiles equal to 3. This means that wherever a bone has more than three consecutive bad cross sections, it is split at that point, producing two new bones. This is a rather conservative approach, and is designed to ensure that in case there is a problem with the fitting due to some unexpected effect, a filament will not be cut without reason. The output of the algorithm is a list of bones. Which of these structures are filaments? Filaments should be elongated structures, i.e. have a large length to width ratio. Again conservatively, we set the lower limit of this ratio to 3. This criterion is the last step in deciding whether skeleton bones can be called `filaments'.

\section{Results} \label{sec:results}

\begin{figure*}
 \centering
  \includegraphics[scale=1]{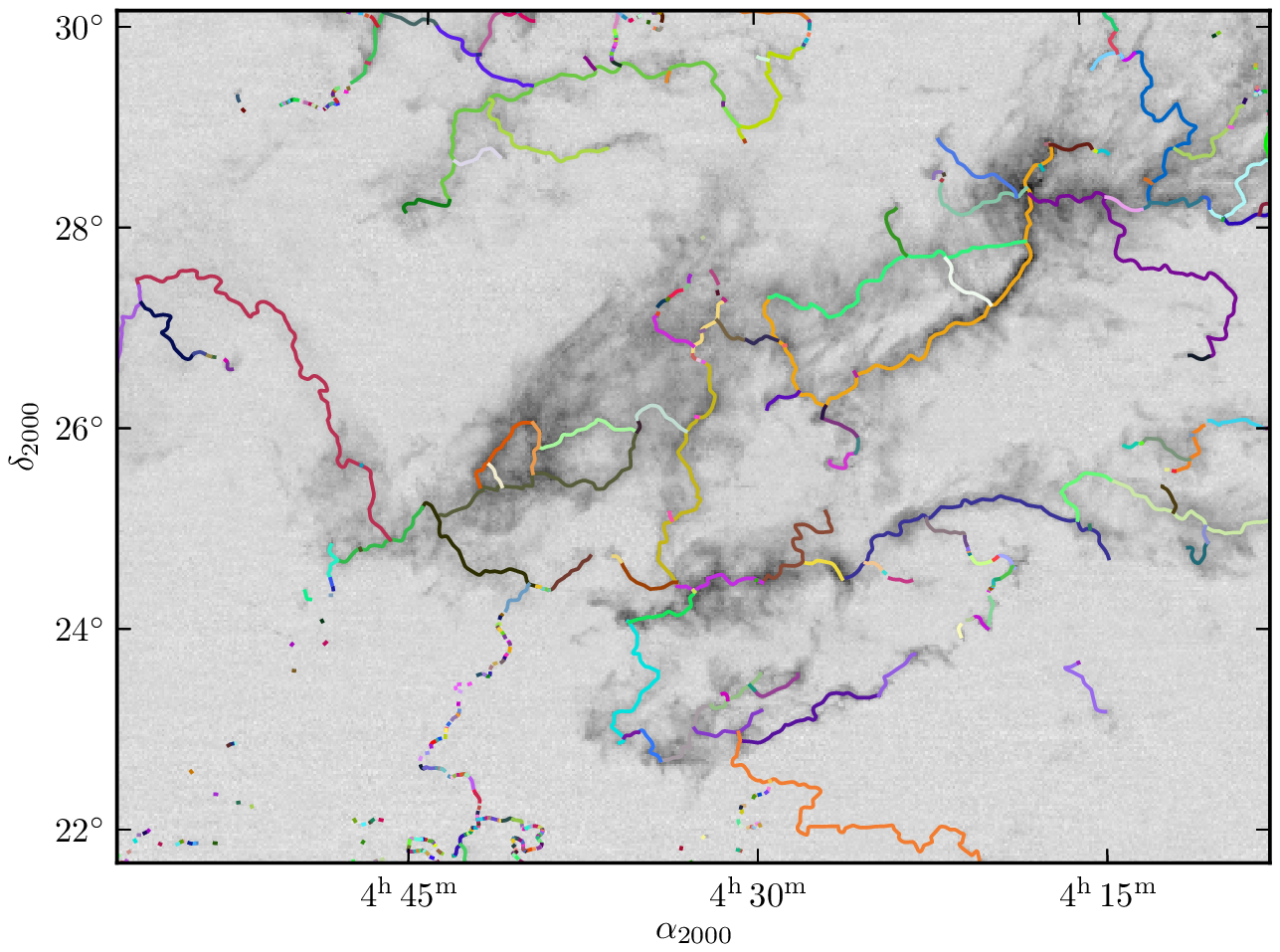}
\end{figure*}
\begin{figure*}
 \centering
  \includegraphics[scale=1]{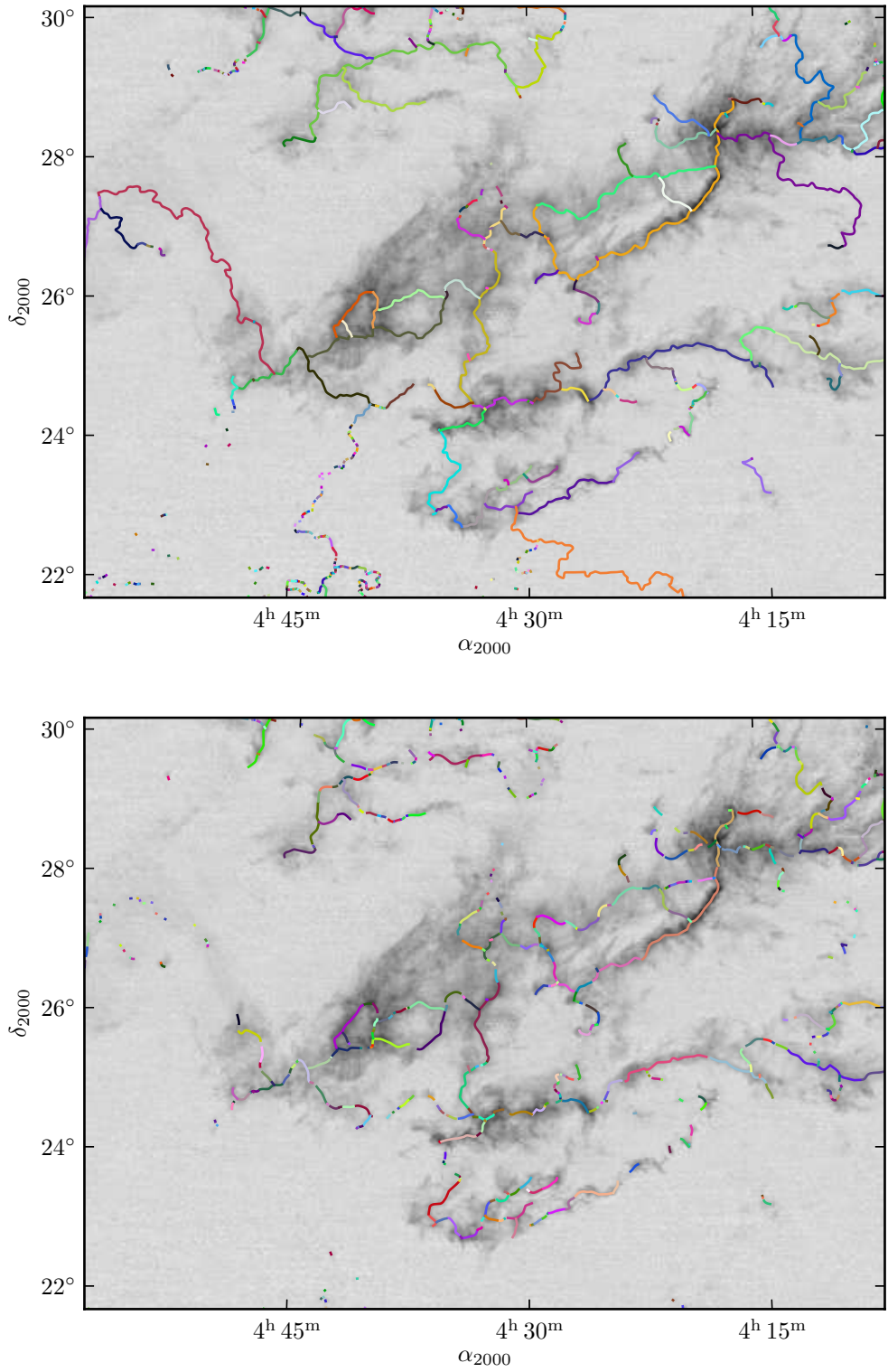} 
\caption{$\rm^{13}CO$ integrated intensity map overplotted with skeletons. Colors show different bones. Top: DisPerSe skeleton with low robustness threshold. Bottom: Post-processed skeleton. See figure \ref{fig:integrated13co} for intensity values.}
\label{fig:13co-skels}
\end{figure*}

\subsection{Filaments in integrated emission}
In this section we present the results of the analysis of the $\rm^{13}CO$ emission map, integrated over the velocity range $0.25$ km/s $- 9.8$ km/s. We ran DisPerSe directly on this data in order to extract the topological skeleton. As explained in section \ref{ssec:Disperse}, there is no single skeleton for a given data set. The DisPerSe parameters need to be adjusted, and the acceptability of a skeleton is largely in the eye of the beholder. Being unable to identify an appropriate threshold in advance, we performed a parameter study and analyzed various skeletons. We present a representative skeleton of the $\rm^{13}CO$ map in figure \ref{fig:13co-skels}. The $\rm^{13}CO$ velocity integrated emission map is overplotted with various colored segments: the `bones' of the skeleton found by DisPerSe (top) and of the profile filtering process (bottom). 

In the presented skeleton, the selected persistence threshold is relatively high, because otherwise the network is too complicated to assess. Robustness is low so that longer structures are traced by the skeleton. This choice of preferential detection of elongated structures is efficient for the most intense areas of the map. To ensure that we do not fail to find lower intensity filaments, we isolated 6 sub-regions of the map and processed them individually. The selection of regions was based on those defined by \cite{Kirk2013} and done so as to aid the analysis of areas of similar intensity. An effort was made to create skeletons with structures as long as possible, in all regions. We choose to be conservative in the selection of the persistence and robustness parameters of DisPerSe so as to identify any filaments in the map, provided they exist. The assembling of skeleton bones was done in a way that allows structures of low significance to remain in the skeleton, but these are later on discarded by our post processing algorithm.

\begin{figure*}
 \centering
\includegraphics[scale=1]{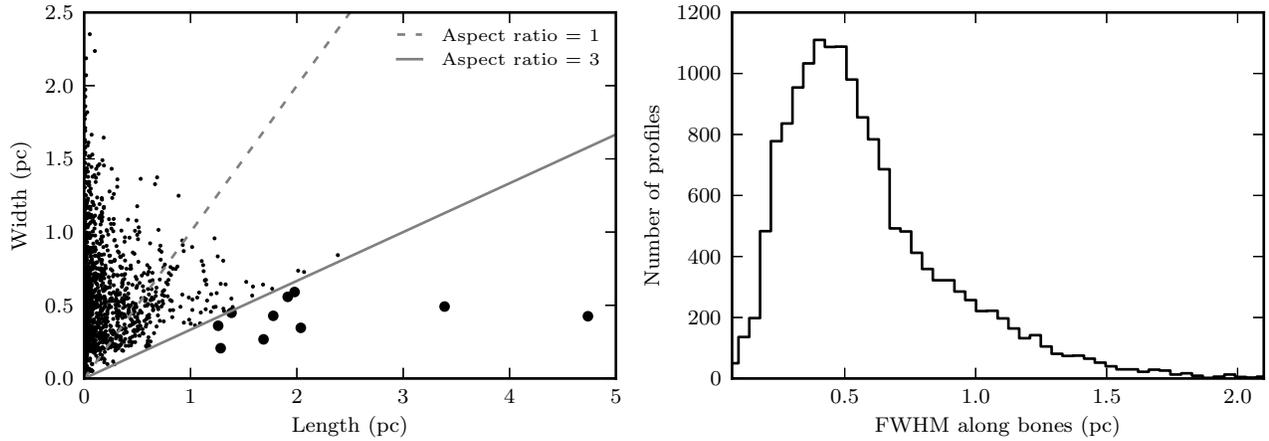}
\caption{Left: Comparison of widths and lengths of bones in the final skeleton (points). The gray dashed and solid lines denote aspect ratios of 1 and 3, respectively. Ten structures qualify as filaments (large dots). Right: Distribution of FWHM of all cross sections across bones in the post-processed skeleton.}
\label{fig:integrated-distros}
\end{figure*}

The first step of the analysis of the bones in the DisPerSe skeleton is to run them through the profiling algorithm. Each cross section of every bone is assigned a width (see section \ref{ssec:profile-filtering}). The right panel of figure \ref{fig:integrated-distros} shows the distribution of cross section widths of all the bones in the post-processed skeleton. A peak is clearly present at around 0.5 pc. The left panel shows the relation between length and width of all bones in the skeleton. Adopting a very conservative approach, we define a structure with aspect ratio $\rm>$ 3:1 as filamentary (provided it has passed all previous criteria). The gray dashed and solid lines denote aspect ratios of 1 and 3 respectively. Bones that pass the ratio threshold of 3 are denoted by bold dots. In the entire map, only 10 filaments are detected. Their properties are presented in section \ref{ssec:filaments}. It is evident, both from this diagram and the skeleton in figure \ref{fig:13co-skels}, that most identified structures have small lengths and a large scatter in widths. These create the tail of the FWHM distribution. Looking at longer structures, we see that the scatter in widths is reduced, with most profile widths around $\rm \approx 0.4 - 0.5$ pc. These widths are generally much larger than the `characteristic width' (0.1 pc) of filaments found in dust emission \cite{arzoumanian2011}.

\subsection{Properties of identified filaments}
\label{ssec:filaments}
\begin{figure}
 \centering
\includegraphics[scale=1]{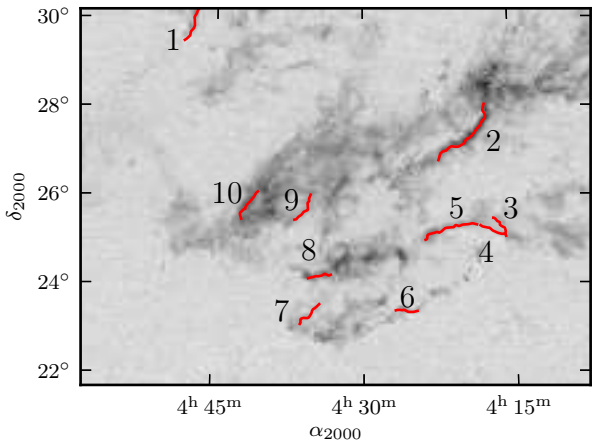}
\caption{Skeletons of the ten filaments found in the image of integrated $\rm^{13}CO$ intensity. See figure \ref{fig:integrated13co} for intensity values.}
 \label{fig:allfils}
\end{figure}

Analysis of the integrated intensity map as well as the separate regions reveals the existence of \textit{10 filaments}. Amongst them are well-studied filaments such as L1495/B213, L1506 (e.g. Pagani et al. 2010), TMC1 (e.g. Malinen et al. 2012). The skeletons of these filaments detected in the integrated emission are overplotted on the map in figure \ref{fig:allfils}.  

\begin{figure*}
 \centering
\includegraphics[scale=1]{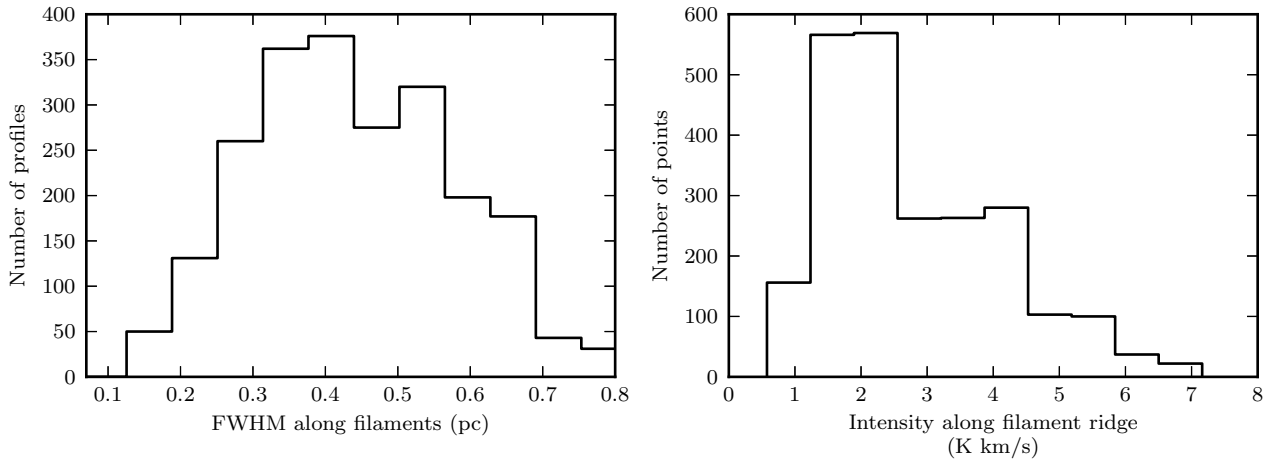}
\caption{Left: Distribution of FWHM of all of the filament profiles. Right: Distribution of intensities along filament ridges.}
\label{fig:integrated-distros2}
\end{figure*}

The distribution of FWHM of all profiles along the filaments can be seen in the left panel of figure \ref{fig:integrated-distros2}. The majority of profiles have a FWHM between 0.2 and 0.6 pc. The peak of the distribution is close to 0.4 pc and has a standard deviation of 0.2 pc. The right panel shows the distribution of intensities along filament ridges, which is peaked at 1.5 K km/s.

\begin{figure}
 \centering
\includegraphics[scale=1]{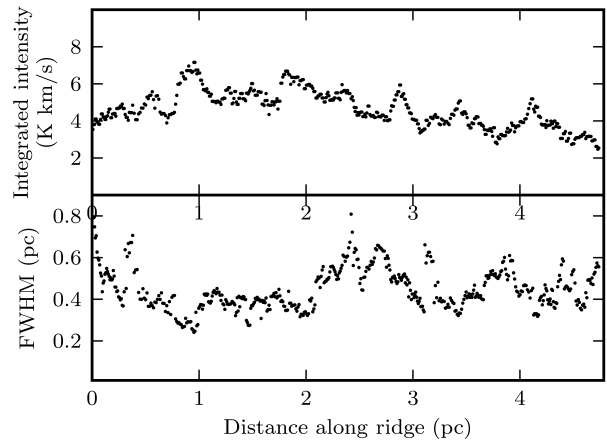}
\caption{Top: intensity of filament 2 along the ridge. Bottom: FWHM of filament along the ridge.}
 \label{fig:meanprofridge}
\end{figure}

Filament 2, which is the most prominent structure in the entire map, is the L1495/B213 filament. The intensity along its ridge (figure \ref{fig:meanprofridge}, top) presents an oscillatory pattern. The FWHM along the ridge (figure \ref{fig:meanprofridge}, bottom) is also nonuniform and has a scatter of $\rm \approx 0.2$ pc. The intensity and FWHM along the ridge do not present any significant correlation in any of the filaments detected. The variation of intensity seen here is similar to that found by Hacar \& Tafalla (2011) in considerably smaller filaments (length $\approx 0.5$ pc) containing cores in the L1517 region of the Taurus cloud. 

\begin{figure*}
 \centering
\includegraphics[scale=1]{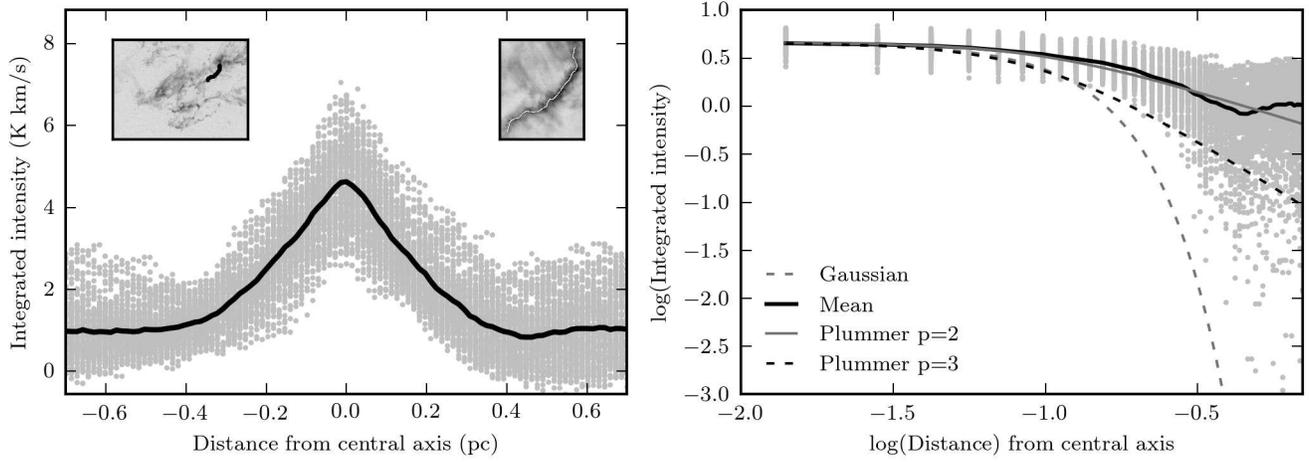}
\caption{Left: Mean profile of filament 2. Gray dots are all the points of the cross sections taken along the filament. Their mean is traced by the black line. Plots in the upper corners show the position of the filament on the map (left: skeleton on the entire map, right: zoomed in area). Right: Mean profile of filament 2 displayed in log-log scale. The dashed gray line is a Gaussian fit to the mean profile (solid black line), the black dashed line is a Plummer function with p=3 and the gray solid line is the same with p=2. The Plummer functions have the same $\rm R_{flat}$.}
 \label{fig:meanprof2-log}
\end{figure*}

\begin{figure*}
 \centering
\includegraphics[scale=1]{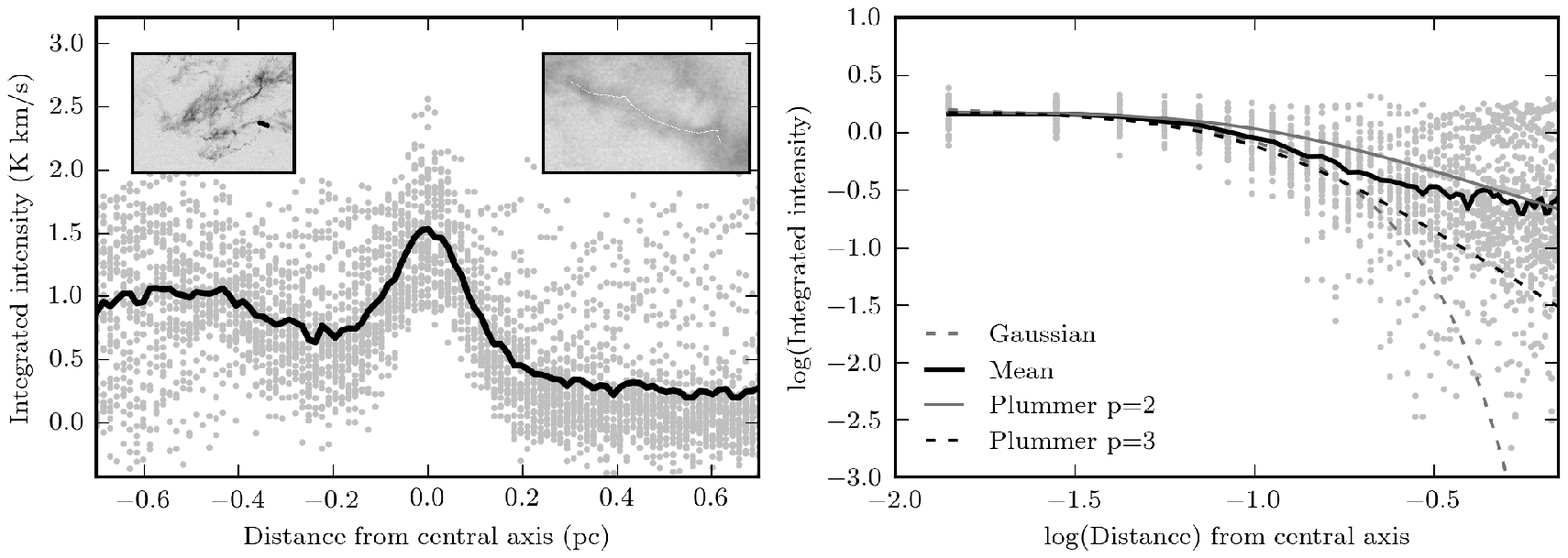}
\caption{Mean profile of filament 4 (as in figure \ref{fig:meanprof2-log}).}
 \label{fig:meanprof4-log}
\end{figure*}

Figure \ref{fig:meanprof2-log} (left) displays the mean profile of filament 2. Gray points are the values of all the profiles along the filament. Their mean is traced by the black line. Embedded plots in the upper left and right corners show the position of the filament in the map and a zoomed-in version of its skeleton plotted on the intensity data, respectively. Figure \ref{fig:meanprof2-log} shows a logarithmic plot of the mean profile. Apart from the mean profile which is denoted by the solid black line, a Gaussian (gray dashed), a Plummer function with p = 2 (solid gray) and with p = 3 (black dashed) are also shown. The Plummer function  with p = 2 seems to follow the shape of the mean profile of this filament up to the radius at which background emission is prevalent. However, the mean profile meets the background at much smaller radii than the extents of dust continuum profiles \cite{arzoumanian2011}, limiting our ability to determine the parameters of the power law. Figure \ref{fig:meanprof4-log} shows the mean profile of a fainter filament, number 4 in figure \ref{fig:allfils}. The profile of this filament seen in figure \ref{fig:meanprof4-log} is steeper than the Plummer function with p=2 (solid gray line) but flatter than p=3 (dashed gray line).


Note that these filaments are structures in a 2D projection of the data. They are the result of the integration of a position-position-velocity data cube. This fact is certain to have strong implications on our perception of the existent (or not) structures. On the one hand, structures present in velocity included data can be suppressed due to addition of diffuse emission from different velocities. On the other, filamentary structures may appear as such in the integrated intensity map only due to projection effects (section \ref{sec:effects}).

\subsection{Filament velocity structure}
\label{ssec:velocities}
 
To study the kinematics of the filaments identified in our data, we use the open source software package ROBOSPECT. It is a program designed to automatically measure and deblend line equivalent widths in spectra~\cite{WatersHollek2013}. It has been previously used to analyze emission spectra of the Orion Nebula, among others. ROBOSPECT fits a model spectrum to the data (with minimum user interaction), which provides the $\rm ^{13}CO$ $J = 1 \rightarrow 0$ line centroid velocities.

\begin{figure*}
 \centering
 \includegraphics[scale=1]{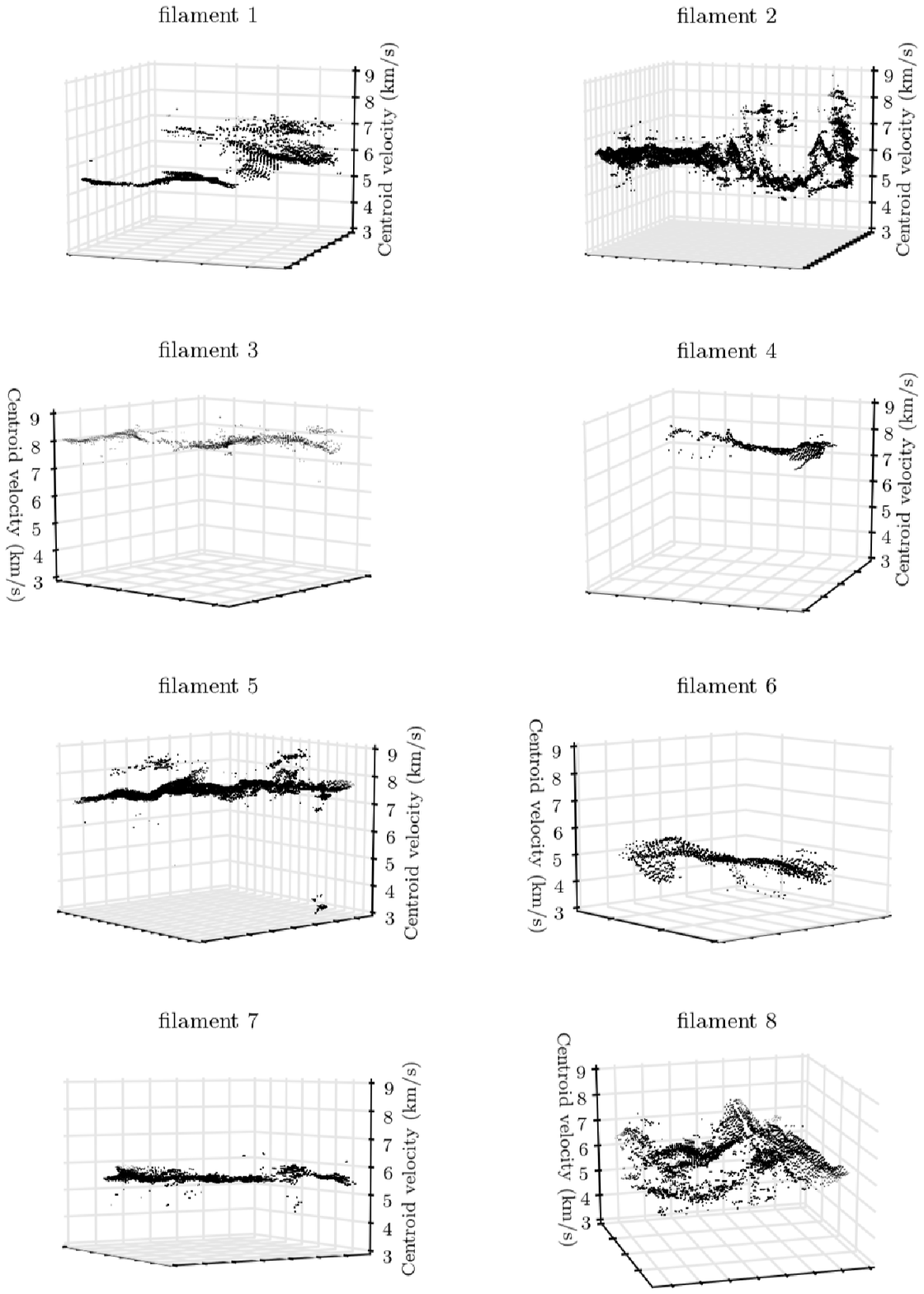}
\caption{PPV diagrams of the 10 filaments. Each dot represents a line peak (in the spectrum fitted by ROBOSPECT). The vertical axis shows the velocity in km/s. Horizontal axes denote distances from the one edge of the filament on the plane of the sky. The grid spacing is 0.5 pc. Points shown are within 0.28 pc of the ridge.}
\label{fig:fils-ppv}
\end{figure*}

The fits reveal the presence of multiple velocity components within individual spectra, consistent with the findings of \cite{hacar2013} for the L1495/B213 filament. A visualization of the velocity information is possible with the position-position-velocity (PPV) diagram, in which points have three coordinates: x and y denote distances on the plane of the sky in parsecs, while the vertical v shows the fitted value of a line peak in km/s. Figure \ref{fig:fils-ppv} presents the structure of the 10 filaments in PPV space, where only pixels along each filament and within 0.28 pc on of the ridge are drawn. From figure \ref{fig:fils-ppv} it is evident that these filaments can be categorized into two groups: those comprised of a narrow range of components (filaments 3, 4 and 7), and those presenting a larger spread in velocity space. Furthermore, the shape of the structures in PPV is not always continuous, with distinct groups of points sometimes present. Caution must be exercised in interpreting structures in these diagrams, as they do not translate uniquely to 3D space \cite{beaumont2013}. 

\begin{figure*}
 \centering
 \includegraphics[scale=1]{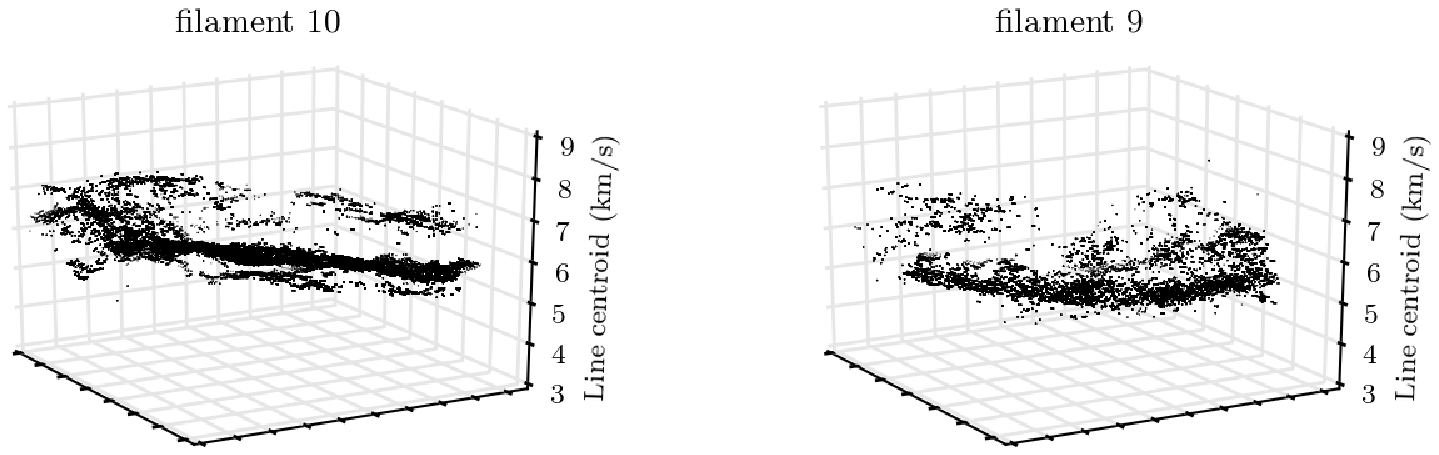}
\caption{As in figure \ref{fig:fils-ppv}.}
\label{fig:fils-ppv1}
\end{figure*}

\begin{figure*}
 \centering
 \includegraphics[scale=1]{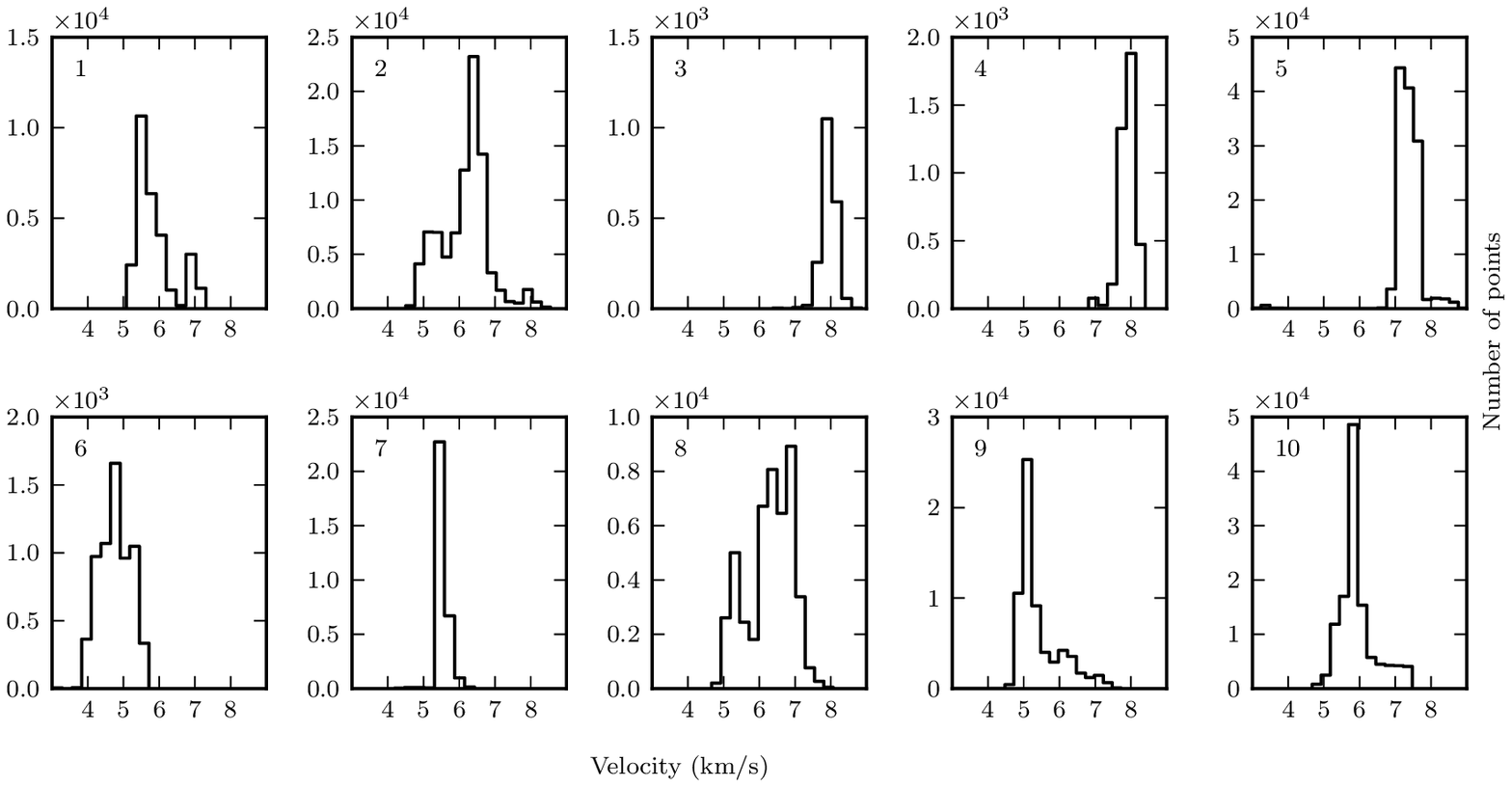}
\caption{Distributions of fitted line peaks for each of the filaments (labeled in the top left corner of each panel). The horizontal axis is the velocity in km/s. }
\label{fig:fils-vhist}
\end{figure*}

Figure \ref{fig:fils-vhist} shows the distribution of velocities of different components found within 0.28 pc of each filament ridge.
These distributions provide a standard deviation, $\sigma _{v}$, of the peak velocities of the different velocity components.

In order to better understand the meaning of the spread of the different velocity components, we use it to calculate the Virial line-mass of each filament. For a long cylinder with negligible external pressure and no magnetic fields, the Virial theorem gives \cite{fiege2000}:
\begin{equation}
\lambda_V = 2 \frac{\langle \sigma_v^2 \rangle}{G},
\label{eqn:virial}
\end{equation}
where $\lambda_V$ is the Virial mass per unit length of the filament and G the gravitational constant. Assuming $\rm ^{13}CO$ is optically thin, the mass per unit length of the emitting material is proportional to the column intensity of the filament. The mass per unit length of the filament is therefore proportional to the product of the mean integrated intensity, $\langle I_f \rangle$, with the square of the filament width, $w_f$,
\begin{equation}
\lambda_f \propto \langle I_f \rangle w_f^2.
\label{eqn:mass-intensity}
\end{equation}
In order to compute $\lambda_f$, we need normalizing values of mass per unit length, intensity and width ($\lambda_n$, $\langle I_n  \rangle$, $w_n$) to produce the following relation: 
\begin{equation}
 \lambda_f = \frac{\lambda_n w_f^2}{\langle I_n  \rangle w_n^2} \langle I_f  \rangle.
\label{eqn:intensity-estim}
\end{equation}
If we assume that dust and $\rm^{13}CO$ emission are proportional to each other, we can use the mass per unit length of filament B211/B213 reported by Palmeirim et al. (2013), $\rm 54 M_\odot/pc$ as well as the values of its mean intensity measured in our map and width found by the fitting process. 

Table \ref{tab:vesc} shows the line-of-sight dimensions, mean intensities, spread of velocity components as well as the corresponding Virial line-mass and estimated line-mass from equation \ref{eqn:intensity-estim}. Faint filaments have substantially fewer velocity components than bright ones, as expected since areas with emission in many velocity channels appear brighter when integrated. The last two columns of table \ref{tab:vesc} show that the Virial line-mass is much larger than the estimated line-mass in most cases. The relative error in mean intensity comes from the statistical spread along the ridge of the filament and is approximately 25\% for all filaments. The width of each filament along its length varies within 0.2 pc of the mean value (relative error $30 - 100$\%) and this variation dominates the uncertainty of this quantity. Validation of our code has shown that the error in the fitting of widths is about 25\% in ideal test maps of smooth constant-width filaments. Palmeirim et al. (2013) place an uncertainty of a factor of 2 on the value of $\lambda_n$. Propagation of errors results in an uncertainty of a factor of 2 for all filament line-masses estimated using equation \ref{eqn:intensity-estim}. Nevertheless, the Virial line-mass exceeds these values by more than a factor of 2 in most cases. Furthermore, due to depletion of CO (and self-absorption, if any) the observed intensity of the main filament is lower than what would correspond to the measured mass. Therefore, from equation \ref{eqn:intensity-estim} it follows that we overestimate the mass of the other filaments which are less intense (and less dense) and do not suffer from these problems. It is thus likely that all filaments have Virial line-masses significantly larger than their estimated line-mass and therefore are probably gravitationally unbound.

A rough estimate of the evolution of the shape of each filamentary structure can be obtained by assuming that it is initially as thick in the line-of-sight direction as the width found on the plane of the sky (cylinder). We calculate the change in size of its line-of-sight dimension implied by the velocity spread after 1 Myr, which is the absolute minimum time interval over which cores are expected to form and is equal to the free fall time for a spherical structure with number density typical for molecular clouds ($\rm n=1000 \, cm^{-3}$)
\begin{equation}
\Delta L_{los} \approx 2\sigma_v t.
\label{eqn:los}
\end{equation} 
The values of the line of sight dimension after this time interval for each filament, shown in table \ref{tab:vesc}, are comparable to the largest projected dimension (length). Thus, the detected filaments may be transient formations on the plane of the sky unless confined by an external pressure component.

\section{Filaments in velocity slices}
\label{sec:slices}

\begin{table*}
\centering
  \begin{tabular}{|c|c|c|c|c|c|c|c|}
    \hline\hline
    Filament  &  Length  & Width &  los size (pc)     & $ \langle I  \rangle$  & $2\sigma_v$ &$\lambda_V$   & $\lambda_f$\\
    number    &    (pc)  &  (pc) &  after 1Myr        &   (K km/s)             &  (km/s)    & ($M_\odot/pc$) & ($M_\odot/pc$)  \\
    \hline
    1         & 2.0      & 0.4   & 1.4                & 2.2                    & 1.1        &  130           & 17         \\
    \hline
    2         & 4.7      & 0.4   & 1.8                & 4.6                    & 1.3        &  205           & 54         \\
    \hline
    3         & 1.3      & 0.4   & 0.7                & 1.6                    & 0.4        &  16            & 13         \\
    \hline 
    4         & 1.7      & 0.3   & 0.7                & 1.5                    & 0.5        &  25            & 7          \\
    \hline
    5         & 3.4      & 0.5   & 1.5                & 2.2                    & 1.0        &  106           & 34         \\
    \hline
    6         & 1.3      & 0.2   & 1.1                & 2.0                    & 0.9        &  84            & 6          \\
    \hline
    7         & 1.8      & 0.4   & 0.7                & 1.4                    & 0.3        &  10            & 17         \\
    \hline
    8         & 1.4      & 0.5   & 1.7                & 3.4                    & 1.3        &  181           & 45         \\
    \hline
    9         & 1.9      & 0.6   & 1.8                & 2.1                    & 1.2        &  174           & 43         \\
    \hline 
    10        & 2.0      & 0.6   & 1.7                & 3.8                    & 1.1        &  131           & 85         \\
    \end{tabular}
    \caption{Filament dimensions on the plane of the sky (length, width), inferred (based on velocity spread) sizes along the line of sight after 1 Myr, mean intensities in the integrated intensity map, full spreads of velocity distributions (2$\sigma_v$), Virial line-masses from equation \ref{eqn:virial} ($\lambda_V$) and estimated line-masses from equation \ref{eqn:intensity-estim}($\lambda_f$).}
\label{tab:vesc}
\end{table*}

We performed a slicing of the data cube in order to search for structures in velocity channels (a method used by e.g. Nagahama et al., 1998). We created 14 channel maps of 0.532 km/s width starting from 2 and ending at 9 km/s. We performed the analysis of these slices using the DisPerSe and post-processing algorithms which resulted in identification of a total of 143 structures which satisfied our criteria and could be deemed `filamentary'.  

\begin{figure*}
 \centering
 \includegraphics[scale=1]{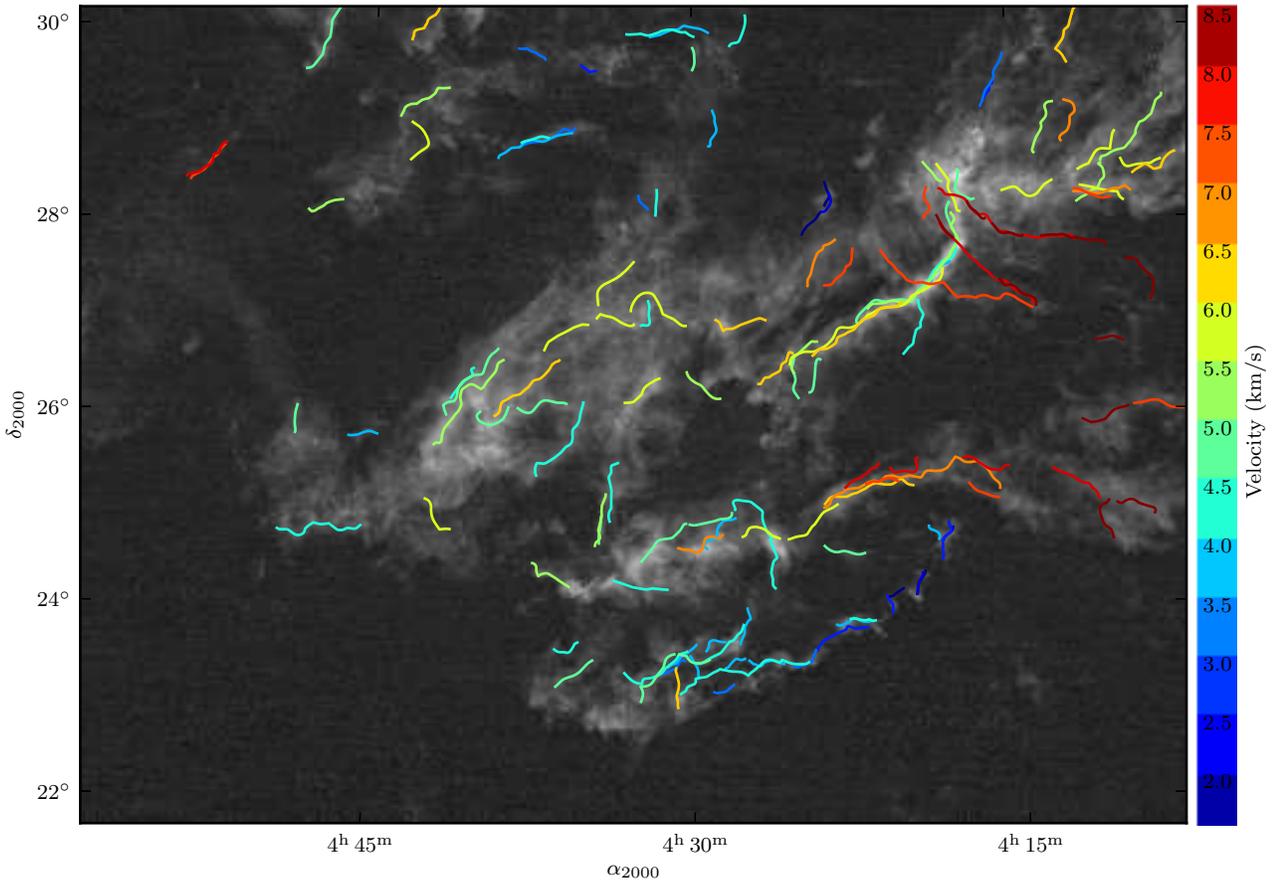}
\caption{Filaments in velocity slices plotted over the 0.25 km/s to 9.8 km/s integrated intensity map. Colors denote the minimum value of the velocity in each slice, and filaments belonging to a specific slice are colored according to the colorbar on the right.}
\label{fig:coloredvfils}
\end{figure*}

Figure \ref{fig:coloredvfils} shows all the filaments found in velocity slices plotted over the integrated intensity map of 0.25 km/s to 9.8 km/s. Filaments are colored according to the lower bound of the velocity interval in which they were identified. Filaments at low velocities reside in the upper and lower parts of the map, whereas those at intermediate velocities occupy the main area of the cloud. Furthermore, those at the highest velocities are situated to the far right of the map. 

The most intense filamentary structures in the integrated intensity map, L1495/B213 and L1506 are comprised of multiple velocity channel map filaments. Interestingly, there are high velocity filaments that appear to be perpendicular to the intermediate velocity L1495/B213 filament(s). These are part of the bubble-like structures found by Chapman et al. (2010), figure 10. Palmeirim et al. (2013) found much shorter low density striations perpendicular to the main filament in the same data. The bright red colored filament crossing B213/L1495 at $\alpha = 4^h20^m00^s$, $\rm \delta = 27^o10'00''$ coincides spatially with the majority of these striations. Previous studies of this particular region \cite{hacar2013} in a similar density tracer, $\rm C^{18}O$, have found a multitude of almost parallel filamentary structures (as is the case in our data) comprising the integrated emission filament. Their approach of searching for structures in PPV space yields shorter structures than the ones found in this slicing analysis. This is most likely due to the difference in techniques, possibly aided by the fact that $\rm ^{13}CO$ traces slightly lower density gas. However, the presence of the high velocity perpendicular structures a few parsecs in length has not been identified previously due to lack of field coverage and possibly difference in density tracers.

\begin{figure}
 \centering
 \includegraphics[scale=1]{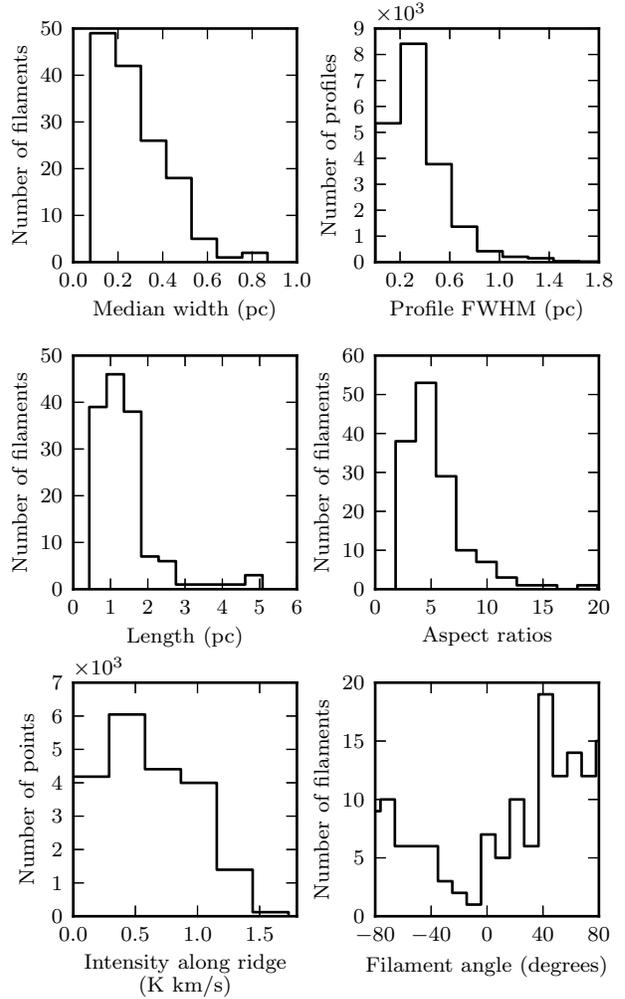}
\caption{Distributions of properties of all filaments in velocity slices. Top left: median of profile FWHMs of each filament. Top right: all profile FWHMs. Middle left: lengths. Middle right: aspect ratios. Bottom left: Distribution of intensities along filament ridges. Bottom right: Distribution of filament angles with respect to line of constant RA. Angles are found by fitting a line to each filament.}
\label{fig:lengths}
\end{figure}

The properties of the filaments found in velocity slices are shown in figure \ref{fig:lengths}. The top panels depict the distribution of median widths (left) and profile widths (right) which exhibit a peak at 0.2 pc, one half of the value found for filaments in the integrated emission (figure \ref{fig:integrated-distros2}). Most filaments have a length less than 2 pc, seen in the distribution on the lower left panel. The aspect ratios of these filaments are clustered around 5, with less than 10 filaments having a ratio of 10 or more (figure \ref{fig:lengths}, middle right panel).

We also examine the orientation of the filaments to search for possible patterns. An approximate value of the angle of each filament with respect to a line of constant RA is obtained by a linear regression. In most cases, this linear approximation is satisfactory for our purpose. The distribution of resulting angles is shown in figure \ref{fig:lengths} (bottom right). The peak at 40 degrees is to be compared to the angle of the L1495/B213 filament which is 60 degrees, as is the mean orientation of the cloud.

The distribution of intensities along the ridge of these filaments, seen in the bottom left panel of figure \ref{fig:lengths}, is much narrower than that of integrated intensity filaments (notice change in horizontal axis) and situated at much lower values.

\begin{figure}
 \centering
\includegraphics[scale=1]{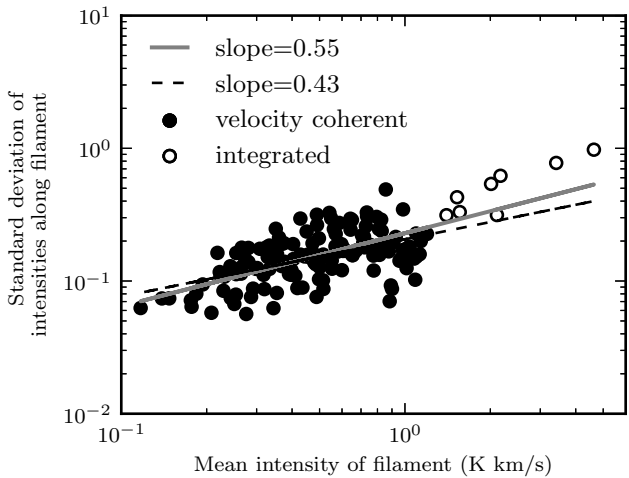}
\caption{Standard deviation of intensity along filaments as a function of intensity (logarithmic). Filled circles correspond to filaments found in velocity channel maps while empty circles are integrated intensity filaments. The gray and black lines are linear fits to the filled dots and all dots respectively.}
\label{fig:intensities-scatter}
\end{figure}

Figure \ref{fig:intensities-scatter} shows the standard deviation of the intensity along the ridge versus mean intensity (with logarithmic axes). Filled dots are filaments found in velocity channel maps while empty dots are integrated intensity filaments. The gray line is a fit to all filaments and the black line only to those in velocity slices. A correlation is not surprising since in faint filaments variations in intensity are bounded by the proximity to the background level. Whereas in intense filaments there is more room for the intensity to vary. The slope of $\approx 0.5$ though, is close to that expected from a Poisson distribution and may indicate that even the filaments found in velocity channel maps have a random origin. 

The existence of hundreds of filaments in the velocity channel maps, contrary to that of an order of magnitude fewer in the 0.25 km/s to 9.8 km/s integrated intensity map indicates that the omission of velocity information greatly affects the result of analyses of filamentary structures.

\section{Projection effects}
\label{sec:effects}

\begin{figure*}
 \centering
 \includegraphics[scale=1]{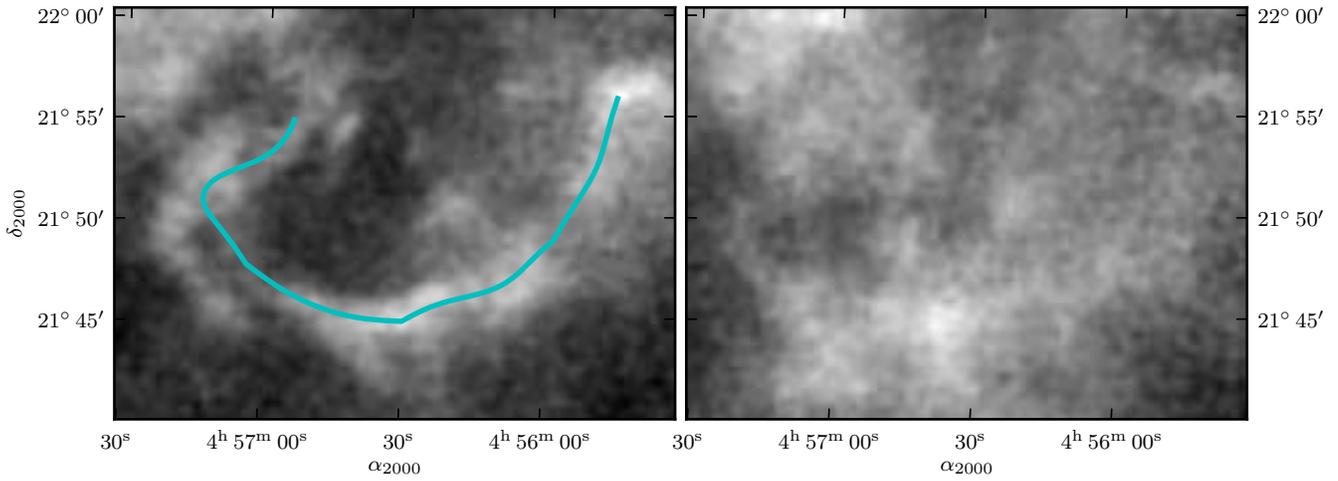}
\caption{Left: The skeleton of a filament overplotted on the 5 km/s $-5.5$ km/s slice. Right: the same region as viewed in the integrated intensity map (0.25 km/s to 9.8 km/s).}
\label{fig:cloaked2}
\end{figure*}

In this section we discuss two effects occurring in our data due to velocity confusion that affect the identification of filaments.

Throughout the data cube, various regions on the plane of the sky emit radiation that does not form any localized shape in the scales associated with filamentary sizes. Within the velocity-integrated intensity images, this diffuse component can mask or hide filamentary structures, whose emissions may be limited to a smaller velocity range. Figure \ref{fig:cloaked2} presents such a case in our own data. In the left panel a filament identified by our analysis is shown in blue, over a small region of the map in the velocity range 5 km/s to 5.5 km/s. The structure is clearly visible. However, the same region in the 0.25 km/s to 9.8 km/s integrated intensity map can be seen on the right and shows no evidence of the underlying structure. This `cloaking' effect could be the reason why such a large fraction of the filaments found in individual velocity slices is obscured in analysis of the integrated emission map.

\begin{figure}
 \centering
 \includegraphics[scale=1]{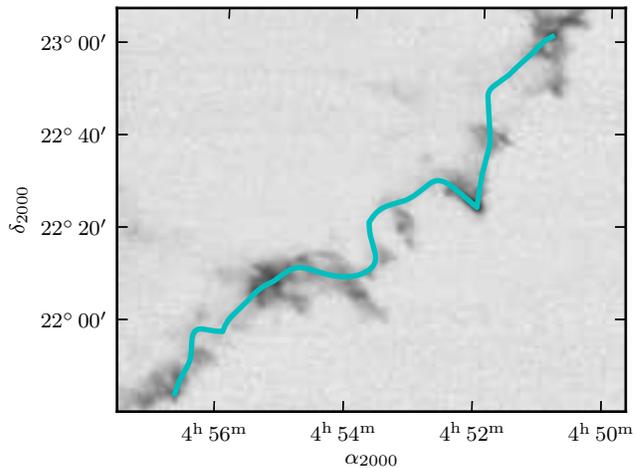}
\caption{Filament skeleton found by DisPerSe in a velocity slice 2 km/s to 4 km/s.}
\label{fig:zsplitfil}
\end{figure}

\begin{figure*}
 \centering
 \includegraphics[scale=1]{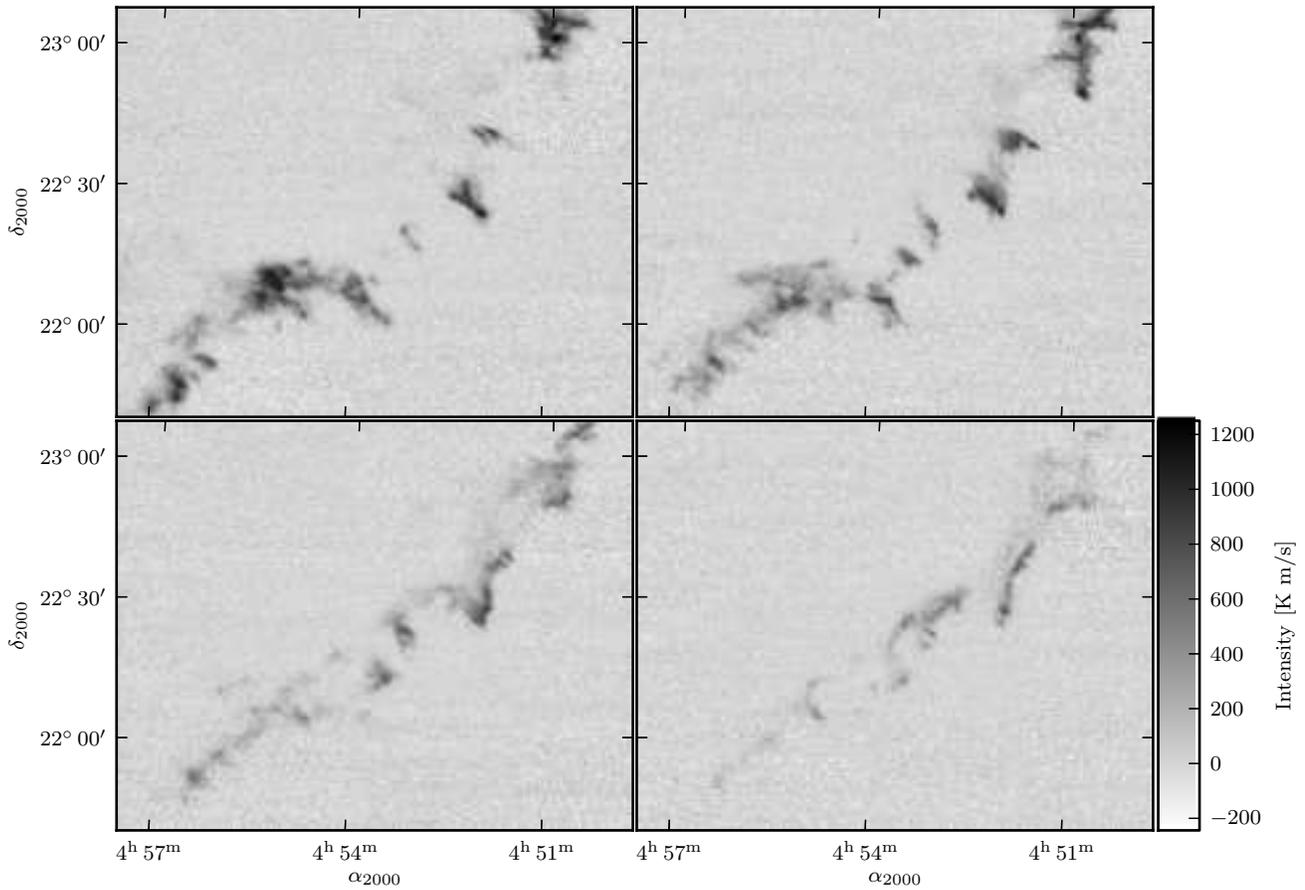}
\caption{Structures found by DisPerSe in velocity slices of 0.5 km/s range. The velocity intervals are: (top left) 3.4 km/s to 3.9 km/s, (top right) 2.9 km/s to 3.4 km/s, (bottom left) 2.4 km/s to 2.9 km/s, (bottom right) 1.8 km/s to 2.4 km/s}
\label{fig:csplitfil}
\end{figure*}

Conversely, localized emission from various velocities can be linked in projection. As a consequence, structures in distinct velocity channels may appear as a single structure in integrated emission, if situated nearby on the plane of the sky. This effect has been identified previously in simulations \cite{juvela2012,moeckel}. An example of this second effect exists in the Taurus data (figures \ref{fig:zsplitfil} and \ref{fig:csplitfil}). Figure \ref{fig:zsplitfil} shows the skeleton of a filament traced by DisPerSe using low thresholds, over the integrated emission in the range 2 km/s to 4 km/s. This filament was identified as an `irregular filament or boundary' by Goldsmith et al. (2008). However, when viewed in velocity channels of 0.5 km/s width, this filament is clearly comprised of discrete, compact structures in velocity similar to cores (figure \ref{fig:csplitfil}).

A large population of velocity-coherent cores spread throughout the Taurus cloud was identified in the same data recently \cite{qian2012}. This fact along with the example presented above, lead us to pursue the idea that projection effects might cause a string of cores to appear as an elongated structure in integrated emission maps. We generated intensity maps from a collection of cores placed randomly in space and studied the parameters that determine whether filamentary structures appear to exist. We also examined the properties of these structures in the maps in which they appear. 

\begin{figure}
 \centering
\includegraphics[scale=1]{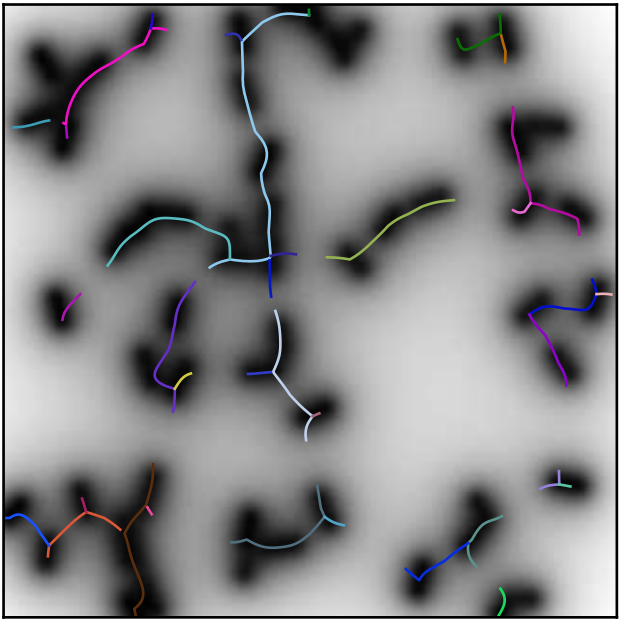}
\caption{Simulated cores with minor axis $\approx$ 14 pixels.}
\label{fig:coremapsfat2}
\end{figure}

Figure \ref{fig:coremapsfat2} shows one of these maps. The images have size 256x256 pixels. Profile-filtered skeletons based on DisPerSe skeletons of relatively low thresholds have been overplotted. The number of cores is the same in all images (60). Cores have random positions and aspect ratios uniformly distributed in the range [1,2]. They are given radial profiles of the form of a Plummer function with p = 2. Their central intensities also vary in a small range [6.5,7.5], similarly to the observational data. Random noise has been added to all images.

\begin{figure}
 \centering
 \includegraphics[scale=1]{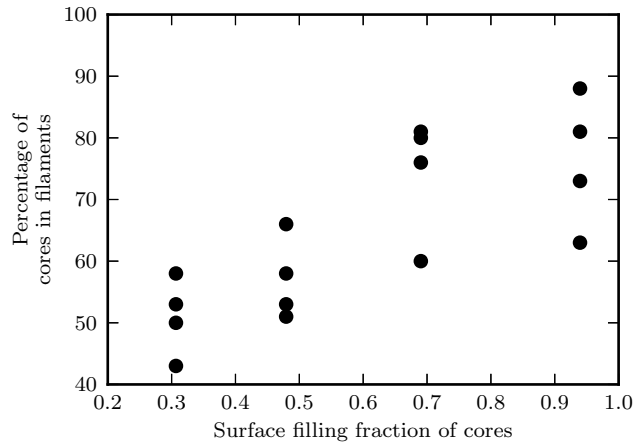}
\caption{Fraction of cores within filaments of DisPerSe skeletons with surface filling fraction.}
\label{fig:sff}
\end{figure}

The parameter that changes from map to map is the minor axis of the cores $R_{min}$, their minimum width which is constant in a single map.  Based on this, we calculate the surface filling fraction, defined as 
\begin{equation}
SFF = \frac{\pi R^2_{min}}{S_{image}}.
\label{eqn:SFF}
\end{equation}
For each value of the size of cores, 4 maps were randomly generated. The approach of keeping the number of cores constant and increasing the SFF by increasing the size of the cores is equivalent to observing different regions of a cloud or of many clouds. Figure \ref{fig:sff} shows the fraction of cores situated in filamentary structures for different surface filling fractions for DisPerSe skeletons. 
As filling fraction increases, the number of cores in structures that can be characterized as elongated, also increases. The spread of the fraction of cores in filaments for the same SFF corresponds to the 4 different random realizations of the maps and reflects the expected statistical uncertainty. This simple test shows that, with sufficient surface filling factor it can be possible that clusters of randomly placed cores on a projected map could be identified as filaments.

\begin{figure*}
 \centering
\includegraphics[scale=1]{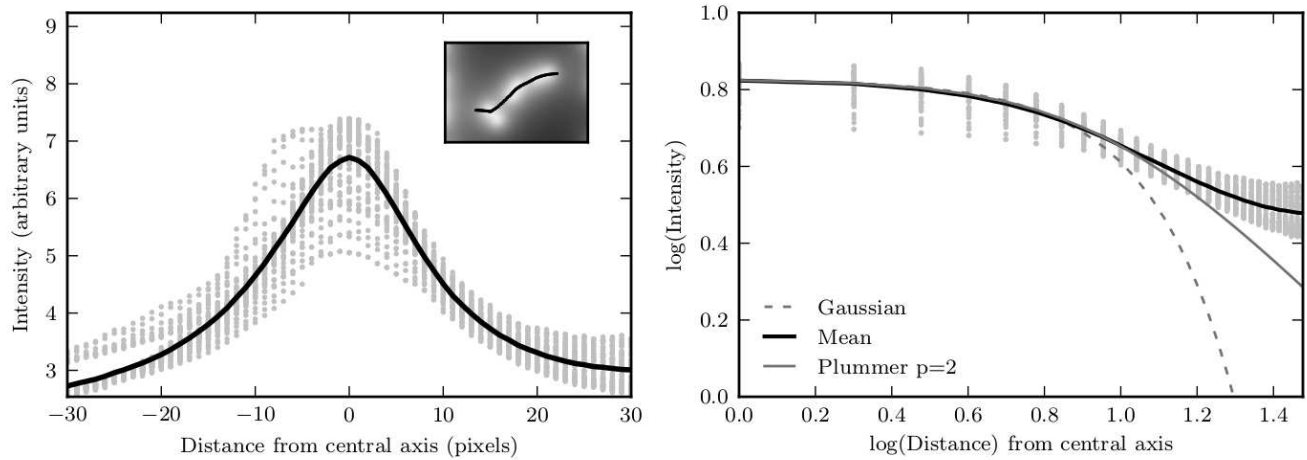}
\caption{Profile of filament comprised of test cores. Left: Linear axes, mean profile is the black line and gray dots are all points in profiles along the filament. Top right corner shows the filament on the map. Right: Logarithmic space, mean is the solid black line, dashed gray line is a Gaussian fit, while solid gray line is a Plummer function with p=2.}
\label{fig:coremapsprof2}
\end{figure*}

In figure \ref{fig:coremapsprof2}, the mean profile of a bone found in the most densely populated map (figure \ref{fig:coremapsfat2}) is plotted in two ways. On the left, the profile is shown with linear axes and its position on the map is indicated in the top right corner. On the right, the mean profile in logarithmic space is shown (black line), with a Gaussian (gray dashed) and a Plummer function (solid gray) fitted to it. It is evident that these properties of cores appearing close in projection do not present any detectable difference compared to the observed properties of filaments found in data. Also, considering that Hacar et al. 2013 found that the clustering of cores on small scales in projection is larger than that expected by a random placement, we expect a more compact positioning of the cores to make the emergence of apparent filaments more likely.

\section{Discussion}
\label{sec:discussion}

The results of the present study reveal the presence of two kinds of filaments in the Taurus molecular cloud seen in $\rm^{13}CO$. Filaments in the integrated emission from 0.25 km/s to 9.8 km/s are very few, present a variety of patterns in PPV space and have a width distribution that peaks around 0.4 pc and has a spread of 0.2 pc. Over 100 filaments are found in velocity channels of 0.5 km/s width. They are velocity-coherent in the sense of being continuous structures within 0.5 km/s bins and their width distribution peaks around 0.25 pc. Integrated intensity filaments have much higher intensities than those found in velocity channels (compare right panel of figure \ref{fig:integrated-distros2} and lower left of figure \ref{fig:lengths}). This is expected, since integration can both enhance/intensify some structures and also suppress others, as explained in section \ref{sec:effects}, resulting in a preference for detection of high intensity filaments. The larger widths of the filaments found in integrated emission compared to those of filaments in velocity channel maps can be attributed to the higher background. This is consistent with validation tests that showed test filaments in background-added maps to be thicker than isolated ones.

Filaments present a variation of widths along their ridge by as much as a few times their mean FWHM, a result consistent with Ysard et al. (2013) who find a variation by a factor of $\approx 4$ along the L1506 filament in Taurus. Most filaments do not present any type of correlation between intensity and FWHM along their ridge.

All filaments in the integrated intensity map, except filament 8, coincide spatially with at least one filament of the channel maps. This suggests that the former are a consequence of the combination of two factors: gas that forms an elongated shape in projection emits in many different velocity channels and diffuse emission is absent or at a level that allows the elongated structures to be discernible. 

The possibility of connection of structures distinct in velocity on the plane of the sky has been addressed by previous theoretical studies (e.g. Beaumont et al., 2013, Moeckel \& Burkert, 2014). An observational example has been provided in section \ref{sec:effects}. The possibility of cores appearing in proximity due to projection could explain the PPV shape of some of the integrated intensity filaments, which consists of distinct groups of points (see figure \ref{fig:fils-ppv} filaments 1, 8). Such an origin could explain the unboundness and short dispersion times calculated in section \ref{ssec:velocities}. Li \& Goldsmith (2012) found that the prominent L1495/B213 filament in Taurus has a high volume density ($\rm10^4 \, cm^{-3}$), by analyzing spectra and excitation conditions of $\rm HC_3N$ along different lines of sight. This result, although presented as favoring a cylinder-like structure, is also consistent with the hypothesis that this filament is comprised of distinct cores overlapping in projection.

The opposite effect, that of blending of coherent filaments with the background, is prevalent in the data. It is the main cause of the large difference in number of filaments found in integrated and velocity channel maps. The fact that filaments are `hidden' in the integrated emission is puzzling when considering the multitude of filaments found in dust continuum emission. Since the dust traces the entire depth of the cloud, and since more diffuse regions outside filaments are warmer than the dense concentrations, the background dust emission should be equally or more prominent than of $\rm^{13}CO$ as the molecular emission is not as sensitive to temperature. However, an important fact concerning CO is that it suffers from freeze-out onto dust grains in the cold, dense regions of the molecular cloud. So it is possible that the densest parts of filaments, that would make them highly contrasted with the background, are not traced by CO.

As with the data of Kirk et al. (2013), we are unable to verify the power law component of the mean radial intensity profiles, as they merge with the background in only several tenths of a parsec. Due to depletion, CO can only trace regions up to a few thousand cm$\rm ^{-3}$. For a characteristic filament width of 0.1 pc \cite{arzoumanian2011} the size of the inner region with uniform density (0.033 pc) implies a volume density of $\rm \approx 3 \cdot 10^5 cm^{-3}$ \cite{tassisyorke}, which is high enough for depletion to take place.

\begin{figure*}
 \centering
 \includegraphics[scale=1]{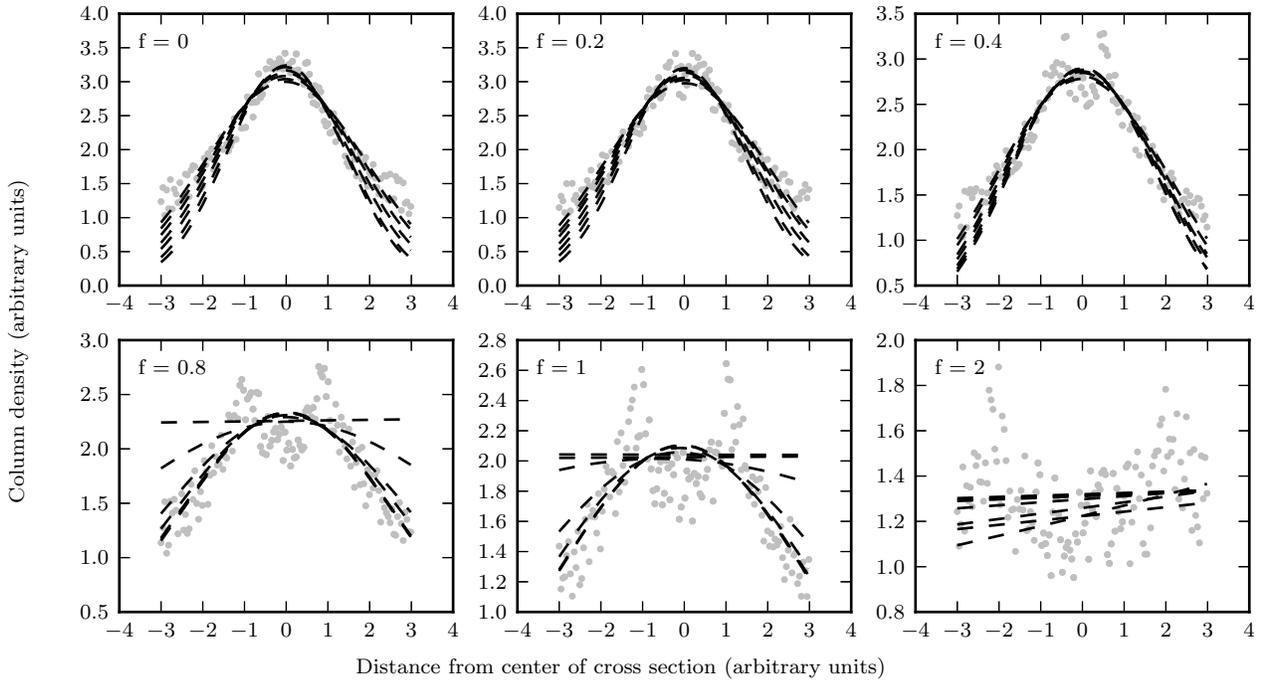}
\caption{Cross section profiles of a simulated isothermal cylinder with CO depletion in a central area with varying radius $\rm r_{d}$. The ratio $\rm f = \frac{r_{d}}{R_{flat}}$ is shown in the top left corner of each panel. Gray dots are the points of the cross section and dashed lines are the Gaussian fits resulting from dynamical fitting. The axes have arbitrary units.}
\label{fig:depletion}
\end{figure*}

Contrary to the Herschel results for dust emission, we do not find a characteristic width either for the filaments identified in integrated emission or for those found in velocity slices. Rather, we find a broad distribution of widths in both cases. In order to determine whether CO depletion can be responsible for this discrepancy, we have roughly estimated its effect on the observed width of a filamentary structure. Under the assumption that the filament is an isothermal cylinder with a Plummer profile with p=2, and that CO is entirely depleted within a central region of radius $\rm r_{d}$ \cite{tafalla2002}, we calculated its column density profile. We then varied $\rm r_{d}$ to simulate an expansion of the depletion region corresponding to different filament ages. Figure \ref{fig:depletion} shows cross section profiles of the column density of the cylinder at different values of the fraction $\rm f = \frac{r_{d}}{R_{flat}}$. If this fraction is less than about 1, the width of a cross section profile is broadened by up to 10\%. Otherwise, the effect of depletion causes the filament to appear as two distinct structures. Therefore, depletion can not in itself explain the significant width difference between the filaments found in this study and those of Herschel column density maps.

However, an effect that may actually have an influence on the shape of the filament profiles is optical depth. If $\rm^{13}CO$ emission becomes optically thick towards the central positions of the filaments, then the spatial profiles become less centrally peaked leading to larger FWHM values. Furthermore, filaments of different density could be affected differently, and this may cause a broadening of their width distribution. As this effect warrants detailed investigation, we plan to address it by comparing the $\rm^{13}CO$ map with Herschel dust continuum data and column density maps derived from the both in a following publication.

Recent publications have discovered velocity gradients along filaments typically of $0.4 - 0.8$ km/s/pc \cite{jimenez2014} and 1.4 km/s/pc \cite{kirkmyers2013}. These are consistent with the filaments found in velocity channel maps, but are much different from what we see in the most velocity-dispersed filaments of the integrated emission map. These parsec-long filaments (1, 2, 7 and 8) are comprised of velocity components that differ by several km/s. The nonlinear gradients of the filaments are consistent with the findings of Kirk et al. (2013) who also find a complex variation of velocities along the filament in the Serpens South region. Signs of continuous longitudinal collapse, such as those found by Peretto et al. (2014) in SDC13 or predicted and observed in S106 by Balsara, Ward-Thompson \& Crutcher (2001), are not apparent in the filaments of the integrated emission map. 

Hacar et al. (2013) reported the existence of sub-filaments in the L1495/B213 filament. Indeed, we find a few filaments in velocity channels extending in the region of the main filament. These filaments are much longer than those found in their study although it is recognized by these authors that the lengths of their structures are over-estimated, as their algorithm attaches different velocity components to find filaments in PPV space. Our choice of bins of 0.5 km/s may cause grouping of multiple short structures into one of larger length. The fact that Hacar et al. (2013) use $\rm C^{18}O$ emission, which traces slightly higher densities, may also result in finding structures of smaller sizes. Unlike these authors, we do not search for filaments directly in PPV space because there is no straightforward way to define a width at every pixel of the PPV cube.

Concerning the paucity of filaments, we should point out that our choice of parameters in the analysis has been conservative. The requirement that filaments have aspect ratios larger than 3 is a very moderate one. Cores are found to have aspect ratios of $1 - 2$ and filaments in the Herschel maps have ratios larger than 5. The same applies for the process of separation of a bone with three consecutive unacceptable profiles. Consequently, we are confident that we found most, if not all, filaments that exist in the integrated intensity map.

\section{Conclusions}

In this paper, we have presented a new method for automatically assessing the topological structures in molecular cloud maps. We use the DisPerSe software to produce topological skeletons and then process its output to obtain a set of astrophysically significant structures.
We implemented the method in searching for filaments in a $\rm100 ~deg^2$ map of the $\rm^{13}CO$ data of the Taurus molecular cloud from the Five College Radio Astronomy Observatory CO Mapping Survey \cite{goldsmith2008,narayanan2008}. 

In the map of integrated emission in the velocity range of 0.25 km/s to 0.98 km/s we find that, only 10 structures of the skeleton are indeed continuous, elongated structures with cross section intensity profiles peaked on the spine of the filament and aspect ratios larger than 3. We also search for such structures in velocity channel maps of 0.5 km/s width in which we find a multitude of filaments. The main results of our analysis are:
\begin{itemize}
\item Filaments in the $\rm^{13}CO$ integrated emission are relatively few. 
\item Intensity profiles along filaments and their corresponding widths have a significant variation.
\item The distribution of all widths along filaments is peaked at 0.4 pc and has a standard deviation of 0.2 pc.
\item Filaments have multiple component velocity structures. The velocity components in each filament have a large enough spread that even if the filament were a cylinder-like structure it would disperse in the line-of-sight dimension in 1 Myr unless confined by an external pressure component.
\item The highest intensity filament presents the largest range of velocity components, their difference being as much as 2 km/s. 
\item More than 100 filaments exist in 0.5 km/s velocity slices of the data cube.
\item The distribution of widths of filaments in the velocity channel maps is peaked at $0.2 - 0.3$ pc. Their angles follow a preferred direction which is approximately that of the cloud. 
\item The intensity along the ridge of filaments in channel maps is lower than that of filaments detected on the integrated emission map. 
\item The spread of the intensity along a filament ridge scales with its mean intensity as $\rm\sigma_I \propto \sqrt{I}$. This is consistent with a stochastic origin of the column density of these structures.
\item Each but one integrated intensity filament coincides spatially with filaments found in channel maps. 
\end{itemize}

We find filaments in velocity channel maps comprising the main L1495/B213 filament at intermediate velocities. Also, we detect 3 parsec-long  filaments, in channel maps, perpendicular to this at high velocities.

The study of structures in integrated intensity maps can be misleading in two ways that have opposite effects. On the one hand, integration may obscure structures that are filamentary in a small number of velocity channels, either due to suppression with comparison to other higher intensity regions in the final map, or the existence of diffuse emission in other velocity channels. On the other hand, structures that are distinct in PPV space may overlap and create the impression of a whole when integrated. We briefly investigated this idea and have seen that a) in maps containing only cores, filamentary structures can be identified by DisPerSe and our profile filtering algorithm, b) properties such as the mean radial profile of these structures are indistinguishable from those of some filaments found in the integrated emission. In conclusion, we find that a collection of cores that are close enough to each other may appear to form filamentary structures in projection. Since cores have an $r^{-2}$ radial profile, the resulting apparent filaments also have the same profile. 

For inquiries about the code please contact the first author (panopg@physics.uoc.gr).

\section*{Acknowledgements}

We are grateful to T. Sousbie for providing the DisPerSe code. We thank V. Charmandaris, N. D. Kylafis, T. Ch. Mouschovias, I. Papadakis, V. Pavlidou, A. Zezas and D. Ward-Thompson for insightful comments which helped us improve the manuscript. We extend our thanks to the authors of the ROBOSPECT code for releasing it for public use.

This research was carried out in part at the Jet Propulsion Laboratory, operated by the California Institute of Technology for NASA.

This research made use of Astropy, a community-developed core Python package for Astronomy \cite{astropy2013}.

K.T. acknowledges support by FP7 through Marie Curie Career Integration Grant PCIG- GA-2011-293531 “SFOnset”.

K.T. and G.V.P. would like to acknowledge partial support from the EU FP7 Grant PIRSES-GA-2012-31578 “EuroCal”.

G. V. P. acknowledges support by the “RoboPol” project, which is implemented under the “ARISTEIA” Action of the “OPERATIONAL PROGRAMME EDUCATION AND
LIFELONG LEARNING” and is co- funded by the European Social Fund (ESF) and Greek National Resources.

\end{document}